%% file: Det_v4.tex
\documentclass[12pt]{article}
\usepackage{jheppub}
\usepackage{tikz}
\usetikzlibrary{positioning}
\usetikzlibrary{calc}
\usepackage{physics}
\usepackage{simplewick}
\usepackage{simpler-wick}

\title{Giant gravitons in twisted holography}
\author[a,b]{Kasia Budzik,}
\author[a]{Davide Gaiotto}
\affiliation[a]{Perimeter Institute for Theoretical Physics, Waterloo, Ontario, Canada N2L 2Y5}
\affiliation[b]{Department of Physics \& Astronomy, University of Waterloo, Waterloo, ON N2L 3G1, Canada}

\emailAdd{kbudzik@perimeterinstitute.ca}
\emailAdd{dgaiotto@perimeterinstitute.ca}

\abstract{We study correlation functions of determinant-like operators in the ``chiral algebra subsector'' of four-dimensional ${\cal N}=4$ gauge theory with $U(N)$ gauge group. We map the large $N$ saddles of the correlation functions to specific semiclassical D-branes in the holographic dual BCOV theory. We present a detailed match of several gauge-theory and BCOV calculations.}

\keywords{Giant gravitons, determinants, twisted holography, chiral algebra, BCOV theory}

\arxivnumber{2106.14859}

\begin{document}
\maketitle
\section{Introduction and conclusions} \label{sec:intro}
The simplest example of Twisted Holography \cite{Gopakumar:1998ki, Costello:2018zrm} is the conjectured equivalence of 
\begin{itemize}
\item The large $N$ 't Hooft expansion for a 2d gauged $\beta \gamma$ system valued in the adjoint representation of $\mathfrak{u}(N)$, aka the ${\cal A}_N$ chiral algebra.
\item The B-model topological string theory (aka BCOV theory) \cite{Bershadsky:1993ta,Bershadsky:1993cx} on an $SL(2,\mathbb{C})$ background with appropriate holographic boundary conditions and coupling $N^{-1}$. 
\end{itemize}
The proposal of \cite{Costello:2018zrm} formulates a holographic dictionary for single-trace local operators whose size remains finite in the large $N$ limit. It includes the match of a large global symmetry algebra which exists on the two sides of the duality at the leading order in $N$. 

In this paper we expand the dictionary to include (sub)determinant operators, whose size scales linearly in $N$. We will match the insertions of such operators to the presence of a B-model D1-brane reaching the boundary of $SL(2,\mathbb{C})$ with specific boundary conditions. 

Correlation functions of (sub)determinant operators admit a rich collection of large $N$ saddles \cite{Jiang:2019xdz}. We will attach to each saddle a {\it spectral curve} in $SL(2,\mathbb{C})$ and identify it with the worldsheet of a dual D1-brane. The explicit identification will allow us to verify a variety of quantitative holographic predictions. 

\subsection{Relation to physical holography}
The Twisted Holography setup we study in this paper is expected to capture a protected subsector of the standard example of holography: the duality between ${\cal N}=4$ $U(N)$ gauge theory and Type IIB string theory on an AdS$_5 \times S^5$ background \cite{Maldacena:1997re,Witten:1998qj}. 

This is manifest on the gauge theory side: the ${\cal A}_N$ chiral algebra and its two-sphere correlation functions coincide with the protected chiral subsector of ${\cal N}=4$ gauge theory introduced by \cite{Beem:2013sza}.\footnote{Torus correlation functions can also be embedded in the gauge theory \cite{Dedushenko:2019yiw,Pan:2019bor}. It would be interesting to study their role in Twisted Holography.} The operators in ${\cal A}_N$ represent specific position-dependent linear combinations of protected operators in the four-dimensional gauge theory. The correlation functions do not depend on the gauge coupling, but have a non-trivial large $N$ genus expansion. 

The relation between the B-model topological string theory on $SL(2,\mathbb{C})$ and Type IIB string theory on AdS$_5 \times S^5$ is less explicit. It is expected to be closely analogous to the original construction relating the B-model to IIB string theory on a self-dual graviphoton background \cite{Bershadsky:1993cx, Antoniadis:1993ze,Ooguri:1999bv}. Both should be special cases of a general construction in twisted supergravity \cite{Costello:2016mgj}. 

The derivation in \cite{Costello:2018zrm} bypassed the twisted supergravity analysis. Instead, 
it employed the near-horizon limit of a stack of $N$ D1-branes in the B-model, in analogy to the original derivation of holography from the near-horizon limit of a stack of $N$ D3-branes in IIB string theory \cite{Maldacena:1997re}.\footnote{Recall that D$n$-branes in the B-model arise from D$(n+2)$-branes in IIB string theory \cite{Ooguri:1999bv}.}

The holographic interpretation of (sub)determinant BPS operators in ${\cal N}=4$ SYM is well understood \cite{McGreevy:2000cw, Balasubramanian:2001nh}. The insertion of a (sub)determinant operator is described holographically as the presence of a ``Giant Graviton'' D3-brane in the bulk, reaching the boundary in a specific manner at the insertion point. 

This entry of the holographic dictionary can be derived by adding a probe D$3'$-brane to the standard near-horizon limit \cite{Jiang:2019xdz}. The D$3'$-brane is taken to be fully transverse to the stack of $N$ D3-branes.  The D$3-$D$3'$ open strings give 0-dimensional fermions coupled to the world-volume theory of the $N$ D3-branes. They can be integrated out to give the (sub)determinant operator insertion. The image of the D$3'$-brane in the near-horizon geometry wraps an $S^3$ inside the transverse $S^5$ and reaches the boundary of AdS$_5$ at the point where the (sub)determinant operator is inserted in the dual gauge theory. 

The derivation can be mimicked in twisted holography. The starting point is a probe D$1$'-brane transverse to the 
stack of $N$ D1-branes. The D$1$'-brane engineers the (sub)determinant operator in the world-volume ${\cal A}_N$ chiral algebra. The image of the D$1$'-brane in the near-horizon geometry wraps a $\mathbb{C}^*$ submanifold in $SL(2,\mathbb{C})$ and reaches the holographic boundary at the point where the (sub)determinant operator is inserted.

\input{GGpic.tex}

In both situations, the holographic dictionary only prescribes that a D-brane should reach the boundary at the 
insertion point with a certain asymptotic shape. The shape of the D-brane in the bulk will be determined dynamically. 
If multiple (sub)determinant insertions are present at the boundary, there may be semiclassical saddles where the 
same bulk D-brane connects all the insertion points, as well as saddles where multiple disconnected D-branes do the job. 

As the 't Hooft coupling dependence drops out in the chiral algebra subsector, we have an exact answer for 
correlation functions of multiple determinants in ${\cal A}_N$ which can be directly compared with a holographic calculation.\footnote{The 't Hooft coupling drops out in a variety of protected subsectors \cite{Corley:2001zk,Berenstein:2004kk}, but these are topological in nature, so the answers only depend on discrete data. 
The chiral algebra subsector allows for a non-trivial (holomorphic) dependence on the positions of the operator insertions,
which must be matched by the dual D-brane geometry.} We identify the holographic dual D-brane as the spectral curve of an auxiliary set of commuting Higgs fields built from the data of the chiral algebra saddle.

The analogous calculation in the physical gauge theory was discussed in \cite{Jiang:2019xdz} and was an important inspiration for this work. In the physical theory, the free answer for generic determinant insertions has large $N$ saddles which can be qualitatively matched to the D-brane saddles. A quantitative comparison, though, requires a full control of the 't Hooft coupling dependence of the answer and is currently out of reach. 

A full lift of our results to the physical holographic duality would require an extra step, which we do not attempt: a match between the holomorphic B-model saddles and explicit supersymmetric D3-brane configurations in AdS$_5 \times S^5$, perhaps along the lines of \cite{Mikhailov:2000ya}. 

\subsection{Saddle equations and spectral curves}
The saddle equations \citep{Jiang:2019xdz} for a correlation function of $k$ (sub)determinants\footnote{A sub-determinant operator is a trace of minors of size smaller than $N$.} in ${\cal A}_N$ take a simple form:
\begin{equation}
[\zeta,\rho] + [\mu, \rho^{-1}] =0 \, ,
\end{equation}
where 
\begin{itemize}
\item $\rho$ is a $k \times k$ matrix whose diagonal elements $m_i$ control the ``length'' of the $i$-th subdeterminant 
operator and whose off-diagonal elements are the variables we are solving for,
\item $\zeta$ is a diagonal matrix whose entries $z_i$ are the positions of the subdeterminant 
operators on the two-sphere,
\item $\mu$ is a diagonal matrix whose entries $u_i$ control the specific linear combination of the adjoint scalars employed in the subdeterminant operator. 
\end{itemize}

Any solution $\rho$ of the saddle point equations allows us to build a family of commuting $k \times k$ matrices
\begin{align}
B(a) &= a \, \mu - \rho \cr
C(a) &= a \, \zeta + \rho^{-1} \cr
D(a) &= a \, \zeta \, \mu + \rho^{-1} \mu - \zeta \, \rho ,
\end{align}
which satisfy 
\begin{equation}
a D(a) - B(a) C(a) = 1 .
\end{equation}
In turn, this defines a spectral curve ${\cal S}_\rho$ in $SL(2,\mathbb{C})$: the collection of points $(a,b,c,d)$ such that $b$,$c$,$d$ are simultaneous eigenvalues of $B(a)$, $C(a)$, $D(a)$. Here we identify $SL(2,\mathbb{C})$ with the locus of $a d - b c =1$ in $\mathbb{C}^4$:
\begin{equation}
g = \begin{pmatrix} a & b \cr c & d \end{pmatrix} .
\end{equation}
Following the holographic dictionary of \cite{Costello:2018zrm}, the spectral curve reaches the 
holographic boundary at $k$ points $z_i$, with an asymptotic shape controlled by $u_i$ and $m_i$ just 
as expected for the support of a holographic dual D1-brane. 

We thus conjecture that a saddle $\rho$ is dual to a B-model 
D1-brane ${\cal B}_\rho$ supported on ${\cal S}_\rho$ in $SL(2,\mathbb{C})$. 
The first test of this conjecture is a comparison of the saddle action $S[\rho]$ on the two sides of the duality. More precisely, we successfully compare the observables $p_i \equiv \frac{\partial S}{\partial m_i}$ conjugate to $m_i$, as well as other derivatives of the saddle action. 

Next, we compare the leading expectation values of single-trace operators in the presence of the collection of (sub)determinant operators. The expectation values are computed from the eigenvalues of a family of auxiliary matrices
\begin{equation}
\frac{\mu -  u}{\zeta - z} \rho^{-1} .
\end{equation}
We recast the chiral algebra calculation as an integral over ${\cal S}_\rho$ of certain 1-forms which we identify with the bulk-to-boundary B-model propagators, completing the holographic match. 

The spectral curve can be connected, but can also consist of multiple components. We analyze the genus 1 part of the large $N$ expansion around a saddle with two disconnected components and compare it with a genus 1 calculation in the B-model. 
We find a non-trivial match, which allows us to probe the Chan-Paton bundle on the holographic branes. The spectral curve comes equipped naturally with a line bundle ${\cal L}_\rho$ given by the common eigenline of $B(a)$, $C(a)$, $D(a)$ corresponding to the $b$,$c$,$d$ eigenvalues. It appears to control the CP bundle for ${\cal B}_\rho$.

\subsection{Determinant modifications and open string boundary conditions}
The holographic dictionary for operator insertions is best understood after the application of a state-operator map. In particular, the determinant operators create a state in the gauge theory which is dual to a state in the string theory where the Giant Graviton D-brane has a certain shape and the D-brane world-volume theory is in its ground state. 

It is also possible to consider operators dual to a Giant Graviton D-brane in an excited state. In the physical theory, they are described by a certain class of modifications of a determinant operator \cite{Balasubramanian:2002sa,Berenstein:2003ah,Balasubramanian:2004nb,Berenstein:2005vf}. 

We formulate such a dictionary for twisted holography and test it with explicit calculations. 
The spectrum of open strings on the B-model brane is well-understood: the worldvolume theory is a $\mathfrak{u}(1)$-gauged $\beta \gamma$ system valued in the normal directions to the brane. The main challenge for us is to produce BRST-closed modifications of 
a determinant operator in ${\cal A}_N$ and to identify BRST-closed modifications which differ by BRST-exact ones. 

Our strategy is to employ the generators of the global symmetry algebra from \cite{Costello:2018zrm} to create the modifications/fluctuations on the two sides of the duality.  Different generators can produce the same open string fluctuations, with specific relative coefficients. On the chiral algebra side the corresponding modifications of determinant operators should satisfy the same relations, up to BRST-exact operators. We test our proposal both by inserting a single modification in the correlation function of multiple determinants and by inserting two modifications in a two-point function.  

\subsection{Structure of the paper}
In section \ref{sec:chiral} we review the relevant aspects of the twisted holography construction and of the construction of determinant operators. In section \ref{sec:det} we derive the large $N$ saddles of a correlation function of (sub)determinant operators and the corresponding spectral curves. We formulate and test the conjecture that the D-branes dual to a given saddle 
are supported on the spectral curve, with a natural choice of the Chan-Paton bundle. In section \ref{sec:corre} we compute the large $N$ expectation value of a single-trace operator in the presence of multiple (sub)determinants. We express the answer in terms of the spectral curves and match it with a B-model calculation. We also match the action of global symmetry generators on the two sides of the correspondence. In section \ref{sec:test} we do a detailed test of our dictionary relating determinant modifications and open string fluctuations on the Giant Graviton. In section \ref{sec:matrix} we review a protected subsector of the chiral algebra which resembles the Dijkgraaf-Vafa setup.

\subsection{Future directions}
The most striking feature of the twisted holography setup is the emergence of geometry from what are essentially free field calculations in the chiral algebra. This reminds us of the protracted efforts to derive a holographic dual description 
of free ${\cal N}=4$ SYM \cite{Gopakumar:2003ns,Gopakumar:2004qb,Gopakumar:2004ys,Gopakumar:2005fx,Aharony:2007fs,Gaberdiel:2021iil,Gaberdiel:2021jrv} and of the Gaussian matrix model \cite{Gopakumar:2011ev,Gopakumar:2012ny} which we encounter in section \ref{sec:matrix}. 

Single-trace operators only probe small fluctuations of the geometry. Determinant operators can create D-branes with non-trivial geometrical shapes. The natural next step is to study operators of size $\sim N^2$, which could 
produce finite modification of the geometry, as in \cite{Lin:2004nb}. Twisted holography could give an explicit, controllable example where multiple geometric saddles contribute to a calculation. 

Another natural objective would be to prove twisted holography at all orders in the large $N$ expansion. Deriving the world-volume theory of Giant Gravitons from the chiral algebra could be a natural first step. Many of the ingredients of the large $N$ expansion 
can be expressed in terms of the spectral curve. It may be possible to recast the expansion in terms of some open string field theory, controlling all string amplitudes which include at least one boundary. 

The twisted holography based on the ${\cal A}_N$ chiral algebra we employed in this paper is one member of a 
large family of examples. Other examples include the chiral algebras associated to ${\cal N}=4$ SYM with other classical gauge groups, as well as ${\cal N}=2$ gauge theories modelled on affine ADE quivers \cite{Costello:2018zrm}. All of these examples have some IIB duals and include a variety of determinant-like operators, often dual to Giant Gravitons wrapping non-trivial cycles in the dual geometries. It may be possible to analyze the corresponding correlation functions with the tools employed in this paper. Correlation functions of determinants and Pfaffians in ${\cal N}=4$ $SO(N)$ gauge theory would be a natural starting point. It would also be interesting to study determinant-like operators 
in less conventional examples, such as class ${\cal S}$ theories with M-theory duals \cite{Gaiotto:2009gz}.  
 
\section{The 2d chiral algebra} \label{sec:chiral}
We will change conventions slightly compared to \cite{Costello:2018zrm} in order to follow the standard 
conventions for a 't Hooft expansion, so that contributions of ribbon diagrams of genus $g$ 
scale as $N^{2-2g}$. 

The relevant chiral algebra ${\cal A}_N$ is defined by a system of gauged symplectic bosons valued in the adjoint representation of $U(N)$. After gauge-fixing, we have an action of the form
\begin{equation}
N \int \,\mathrm{Tr}\left(X \bar \partial Y  + b \bar \partial c \right) .
\end{equation}
Here $X$, $Y$ are a pair of bosonic spinors, aka {\it symplectic bosons}. The ghosts $b$ and $c$ are also valued in the adjoint representation of $U(N)$. The BRST charge has the schematic form 
\begin{equation}
N\oint \, \mathrm{Tr}\left( c [X,Y] +\frac12 b[c,c] \right) .
\end{equation} 
In our calculations we will never actually use the BRST charge or the ghosts. All operators we will employ are built explicitly from a basic set of BRST-closed combinations of the $X$ and $Y$ fields. 

The chiral algebra has an $SL(2)_R$ global symmetry acting on the $(X,Y)$ doublet of symplectic bosons.\footnote{This is a subgroup of the R-symmetry of the physical 4d theory.} Throughout the paper we will employ a very convenient formalism to describe elements of finite-dimensional $SL(2)$ representations: we identify a spin $j$ representation with the space of polynomials of degree $2j$ in some auxiliary variable. The action of $SL(2)$ on the coefficients of a polynomial maps to a fractional linear transformation of the auxiliary variable:
\begin{equation}
p(u) \to (\gamma u + \delta)^{2j} p\left(\frac{\alpha u + \beta}{\gamma u + \delta}\right) .
\end{equation}

In particular, we will employ the linear combination 
\begin{equation}
Z(u;z) \equiv X(z) + u Y(z) 
\end{equation}
and write the symplectic boson OPE as
\begin{equation}
Z(u;z)Z(v;w) \sim \frac{1}{N} \frac{u-v}{z-w} .
\end{equation}

In the large $N$ limit, operators built from a finite number of fields can be expressed as regularized polynomials in single-trace operators. At leading order in $N$, the BRST charge will act separately on each 
single-trace operator in the product. We can thus focus on the BRST cohomology of single-trace operators. 

A word of caution is that the large $N$ expansion does not play well with the full 2d conformal symmetry group: 
most Virasoro descendants of a single-trace operator are multi-trace operators. We will thus simply consider operators with a given scaling dimension and 
their derivatives. 

The single-trace operators
\begin{equation}
A_n(u;z) = N \, \mathrm{Tr}\, Z(u;z)^n 
\end{equation}
are BRST closed. They have $SU(2)_R$ spin $j=\frac{n}{2}$ and scaling dimension $\Delta=\frac{n}{2}$.

The tower of $A_n$ operators does not exhaust the BRST cohomology of single-trace operators. There is a second tower of $D_n$ operators with $SU(2)_R$ spin $j=\frac{n}{2}$ and dimension $\Delta=\frac{n}{2}+2$. These are more complicated to write down explicitly. They can be usefully represented as specific terms in the OPE of operators in the $A$ tower. There are also towers of BRST-closed operators $B_n$ and $C_n$ in ghost number $\pm 1$, but we will not employ them. 

\subsection{The global symmetry algebra}
The chiral algebra correlation functions in the large $N$ limit are invariant under the action of a global symmetry algebra:
the subset of the Fourier modes of single-trace operators which annihilate the vacuum at the origin and at infinity. This algebra can be matched to the algebra of holomorphic vector fields on 
$SL(2,\mathbb{C})$ which preserve the holomorphic three-form $\Omega$. At least at tree level, this is a symmetry for the holographic dual B-model.\footnote{At finite $N$ the Fourier modes of single-trace operators do not form a Lie algebra.} 

The Fourier modes of $A_n(u;z)$ which annihilate the vacuum at the origin and infinity are organized in a multiplet of spin $\frac{n}{2}$ for $SL(2)_R$ and $\frac{n}{2}-1$ for the $SL(2)_L$ global conformal symmetry. We can collect them into a polynomial of degrees $(n,n-2)$ in two auxiliary variables $v$ and $w$:
\begin{equation}
a_{n}(v;w) = \oint \frac{\dd z}{2 \pi i } \, (z-w)^{n-2} A_n(v;z) .
\end{equation}
The Fourier modes of the $D_n(v;z)$ which annihilate the vacuum at the origin and infinity are organized in a multiplet of spin $\frac{n}{2}$ for $SL(2)_R$ and $\frac{n}{2}+1$ for the $SL(2)_L$ global conformal symmetry. We can collect them into a polynomial $d_n(v;w)$ of degree $n$ in $v$ and $n+2$ in $w$.\footnote{There are also fermionic generators associated to the $B_n$ and $C_n$ towers. We will focus on bosonic modes here.}

We can match these Fourier modes with the global holomorphic vector fields on $SL(2,\mathbb{C})$ which preserve the holomorphic three-form\footnote{Written here in the $a \neq 0$ patch of $SL(2,\mathbb{C})$.}
\begin{equation}
\Omega = \frac{\dd a \,\dd b \,\dd c}{a} .
\end{equation}
We can write remarkably compact expressions for the polynomial generating functions of vector fields in the same irreps of $SL(2)_L \times SL(2)_R$. We start with the polynomial $J_2(v)$ collecting the three $SL(2)_R$ generators:
\begin{equation}
J_2(v) = (b - v a)(\partial_a + v \partial_b) +  (d - v c)(\partial_c + v \partial_d) 
\end{equation}
and the polynomial $I_0(w)$ collecting the three $SL(2)_L$ generators:
\begin{equation}
I_0(w) = (c - w a)(\partial_a + w \partial_c) +  (d - w b)(\partial_b + w \partial_d) \, .
\end{equation}
We then match $a_{n}(v;w)$ to some multiple of
\begin{equation} 
J_n(v;w) =(d - v c - w b + v w a)^{n-2} J_2(v)  \label{eq:J}
\end{equation}
and $d_{n}(v;w)$ to some multiple of
\begin{equation}
I_n(v;w) =(d - v c - w b + v w a)^{n} I_0(w) .
\end{equation}
Matching commutators constrains the relation further. In particular, we find that $a_{n}(v;w)$ matches $n J_n(v;w)$.\footnote{
The simplest way to do so is to compare the leading $N$ commutator of $\oint \frac{dz}{2 \pi i} N \Tr X^2 Y(z)$ and $\oint \frac{dz}{2 \pi i}N \Tr X^n (z)$ and the commutator of $b^2 \partial_b - 2 b a \partial_a + b d \partial_d - (a d+b c) \partial_c$ and $n (b^{n-1} \partial_a + b^{n-2} d \partial_c)$. The former gives $n \oint \frac{dz}{2 \pi i} N \Tr X^{n+1} (z)$, the latter gives $n(n+1)(b^{n} \partial_a + b^{n-1} d \partial_c)$. }

It is easy to show that these vectorfields preserve the $a d - b c =1$ locus and thus define vectorfields on $SL(2,\mathbb{C})$.
The invariance of $\Omega$ requires a bit more work. The $SL(2)_R$ variation is 
\begin{equation}
\dd\left(i_{J_2(v)} \Omega \right) = \dd \left(-\frac{\dd c}{a} (b - v a)(\dd b - v \dd a) + (d - v c)\frac{\dd a \dd b}{a} \right)=\frac{\dd a \dd b}{a^2} (-b \dd c + a \dd d) =0 .
\end{equation}
The desired result for $J_n(v;w)$ follows from the observation that $i_{J_2(v)} \Omega$ wedged with 
$\dd (d - v c - w b + v w a)$ vanishes. The analysis for $I_n(v;w)$ is completely analogous. 

There is another useful perspective. Consider the 1-forms
\begin{equation}
\omega_{J_2}(v) = \frac12 \left[ (b - v a)\dd(d - v c) - (d - v c) \dd(b - v a) \right] .
\end{equation}
It is easy to verify that 
\begin{equation}
\dd \omega_{J_2}(v) =\dd(b - v a)\dd(d - v c) = i_{J_2(v)} \Omega. 
\end{equation}
Similarly, define
\begin{equation}
\omega_{J_n}(v;w) =\frac1n (d - v c - w b + v w a)^{n-2} \left[ (b - v a)\dd(d - v c) - (d - v c) \dd(b - v a) \right] .
\end{equation}
We have
\begin{equation}
\dd \omega_{J_n}(v;w) = i_{J_n(v;w)} \Omega. 
\end{equation}
Similarly, we can define 
\begin{equation}
\omega_{I_0}(w) =  \frac12 \left[ (c - w a)\dd(d - w b) - (d - w b) \dd(c - w a) \right]
\end{equation}
and 
\begin{equation}
\omega_{I_n}(v;w) = \frac{1}{n+2}(d - v c - w b + v w a)^{n} \omega_{I_0}(w) .
\end{equation}

These 1-forms play a useful role in describing the action of vector fields on D-branes. 

\subsection{(Sub)Determinant operators}
The determinant operator $\det \, Z(u;z)$ is a BRST-closed operator. We will also consider the ``subdeterminant'' operator, which is a linear combination of operators with different scaling dimensions
\begin{equation}
{\cal D}(m;u;z) \equiv \det \, \left(m+Z(u;z) \right) \, .
\end{equation}
Individual summands in this generating function can be obtained by expanding in powers of $m$.\footnote{We can give the whole generating function a good behaviour under global conformal transformations acting on $z$ and $SU(2)_R$ rotations acting on $u$ by allowing $m$ to transform in the same way as $Z(u;z)$ under fractional linear transformations. }

For concrete calculations, we will describe a determinant operator with the help of some auxiliary fermionic fields:\footnote{These can be interpreted as the D$3$-D$3'$ open string modes. It is also possible to employ a probe D3''-brane, which is only transverse to the stack of $N$ D3-branes in the B-model directions. This would give rise to bosons and inverse determinants.}
\begin{equation}
\det \, \left(m+Z(u;z) \right) = \int \dd \psi  \dd \bar \psi \, e^{\bar \psi \left(m+Z(u;z) \right) \psi} .
\end{equation}

The fermionic language is also useful to describe modifications of a determinant operator. For example, 
something like 
 \begin{equation}
\int \dd \psi \dd \bar \psi  :e^{\bar \psi X \psi}  \bar \psi O \psi:
\end{equation}
defines a modification of the $\det X$ operator where one of the ``$X$'' symbols has been replaced by $O$.

As we will review momentarily, the $\bar \psi O \psi$ insertion will behave as a boundary insertion in the 't Hooft expansion. 
As a consequence of the large $N$ combinatorics, these insertions will behave at the leading order in $N$ 
in a manner similar to single-trace operators: we can place multiple independent insertions in the fermionic integral 
and the BRST charge will act separately on each at the leading order in $N$. It is thus possible to define a BRST cohomology
of modifications of a specific determinant operator.

In practice, it is not immediately obvious how to write down BRST-closed modifications, even at the leading order in $N$. 
Our strategy will be to produce such operators by acting with the global symmetry generators on the (sub)determinant operators. For example, a commutator of the form 
\begin{equation}
\oint \frac{\dd z}{2 \pi i } N \Tr Y^{n+1}(z) \det X(0) = \int \dd \psi \dd \bar \psi  :e^{\bar \psi X(0) \psi}  \bar \psi Y(0)^n \psi: + \cdots
\end{equation}
gives a BRST-closed operator which at the leading order in $N$ contains a single $X \to Y^n$ modification. This decreases the $R$-charge by $\frac{n+1}{2}$ and increases the scaling dimension by $\frac{n-1}{2}$. A similar construction with the component of $D_{n-1}$ of minimal R-charge will produce a modification which decreases the $R$-charge by $\frac{n-1}{2}$ and increases the scaling dimension by $\frac{n+1}{2}$. The fermionic towers of single-trace operators of ghost number $\pm 1$ would similarly give modifications of ghost number $\pm 1$ which increase/decrease the scaling dimension/R-charge  by $\frac{n}{2}$.

We conjecture that these four towers of modifications exhaust the BRST-cohomology
of modifications of $\det X$. This is a non-trivial statement, which is justified by an index calculation \cite{indet} 
and will match the collection of open string fluctuations available on the holographic dual side. 

\section{Correlation functions of determinants} \label{sec:det}
We will now compute a sphere correlation function of subdeterminant operators 
\begin{equation}
\langle {\cal D}(m_1;u_1;z_1) \cdots {\cal D}(m_k;u_k;z_k) \rangle \,  .
\end{equation}
using the techniques of \cite{Jiang:2019xdz}. We will assume the $u_i$ to be distinct for convenience. 

Using the fermionic presentation, this can be written as 
\begin{equation}
\left\langle \prod_i {\cal D}(m_i;u_i;z_i) \right\rangle =\int \prod_i [\dd \psi^i \dd \bar\psi_i]  \left\langle \prod_i  e^{\bar \psi_i \left(m_i+Z(u_i;z_i) \right) \psi^i}\right\rangle 
\end{equation}
and evaluated to 
\begin{equation}
\left\langle \prod_i {\cal D}(m_i;u_i;z_i) \right\rangle =\int [\dd \psi \dd \bar\psi]\,  e^{-\frac{1}{2N}\sum_{i \neq j} \frac{u_i-u_j}{z_i-z_j}\bar\psi_i \psi^j \bar\psi_j \psi^i + \sum_i m_i \bar\psi_i \psi^i} .
\end{equation}
We do an Hubbard-Stratonovich transformation, by introducing auxiliary variables $\rho^i_j$ for $i \neq j$. We also define $\rho^i_i = m_i$. We thus write
\begin{equation}
\left\langle \prod_i {\cal D}(m_i;u_i;z_i) \right\rangle =\frac{1}{Z_\rho} \int [\dd \psi \dd \bar\psi] [\dd \rho] \, e^{\frac{N}{2}\sum_{i \neq j} \frac{z_i-z_j}{u_i-u_j}\rho^i_j \rho^j_i + \sum_{i,j} \rho^i_j \bar\psi_i \psi^j} ,
\end{equation}
where 
\begin{equation}
Z_\rho \equiv \int [\dd \rho] \, e^{\frac{N}{2}\sum_{i \neq j} \frac{z_i-z_j}{u_i-u_j}\rho^i_j \rho^j_i } .
\end{equation}
The $\int [\dd \rho]$ integral only involves the $k(k-1)$ off-diagonal components of $\rho$ and employs whatever contour makes the Gaussian integral converge. 

We can now integrate the fermions away to get the final answer
\begin{equation}
\left\langle \prod_i {\cal D}(m_i;u_i;z_i) \right\rangle =\frac{1}{Z_\rho}\int [\dd \rho] \, e^{\frac{N}{2}\sum_{i \neq j} \frac{z_i-z_j}{u_i-u_j}\rho^i_j \rho^j_i } \left[\det \rho \right]^N .
\end{equation}

This form is suitable for a large $N$ analysis. We can do a saddle-point evaluation of the $\rho$ integral, with action 
\begin{equation}
S[\rho] = \frac12 \sum_{i \neq j} \frac{z_i-z_j}{u_i-u_j}\rho^i_j \rho^j_i + \log \det \rho .
\end{equation}
The classical saddles satisfy the equations
\begin{equation}
\frac{z_i-z_j}{u_i-u_j}\rho^i_j + [\rho^{-1}]^i_j =0, \qquad \qquad i \neq j,
\end{equation}
i.e. in matrix form 
\begin{equation} \label{eq:saddle}
[\zeta,\rho] + [\mu, \rho^{-1}]=0 ,
\end{equation}
where we introduced diagonal $k \times k$ matrices $\zeta$ and $\mu$ with entries $z_i$ and $u_i$. 

From now on, whenever $\rho$ appears outside of an integral it refers to some specific solution of (\ref{eq:saddle}).
Recall that $\rho^i_i = m_i$. It is also useful to define
\begin{equation}
p_i \equiv \frac{\partial S[\rho]}{\partial m_i} = \left[\rho^{-1}\right]^i_i ,
\end{equation}
which helps to probe the dependence of $S[\rho]$ on the choice of saddle. 

\subsection{The Spectral Curve}
As anticipated in the introduction,  the saddle point equations (\ref{eq:saddle}) guarantee the commutativity of a one-parameter family of $k \times k$ matrices
\begin{align} \label{eq:BCD}
B(a) &= a \, \mu - \rho \cr
C(a) &= a \, \zeta + \rho^{-1} \cr
D(a) &= a \, \zeta \, \mu + \rho^{-1} \mu - \zeta \rho ,
\end{align}
which satisfy 
\begin{equation}
a D(a) - B(a) C(a) = 1 .
\end{equation}
This allows us to associate to any saddle $\rho$ a spectral curve ${\cal S}_\rho$ in $SL(2,\mathbb{C})$,
with coordinates $a$,$b$,$c$,$d$ constrained by 
\begin{equation}
a d - b c =1 .
\end{equation}
The spectral curve consists of points $(a,b,c,d)$ such that $b$,$c$,$d$ are simultaneous eigenvalues of $B(a)$, $C(a)$, $D(a)$. 

The spectral curve is non-compact: as $a \to \infty$ we find $k$ branches where $b$, $c$ and $d$ grow linearly with $a$. More precisely, we can expand 
\begin{align}
\frac{b}{a} &=u_i - \frac{m_i}{a} + \cdots\cr
\frac{c}{a} &=z_i + \frac{p_i}{a} + \cdots \, .
\end{align}
We would like to compare these boundary conditions with the holographic boundary conditions expected for 
a dual D1-brane. 

Recall that the holographic boundary conditions of \cite{Costello:2018zrm} control the behaviour of the B-model fields 
as $(a,b,c,d)$ go to infinity with fixed ratios. Single-trace operator insertions at some point $z$ control the behaviour of the fields at fixed $\frac{c}{a}=z$. Analogously, we expect the insertion of (sub)determinant operators to correspond to the 
presence of a D1-brane which asymptotically lies in the $\frac{c}{a}=z$ locus.

\subsection{The asymptotic shape of Giant Gravitons}
The near-horizon analysis of \cite{Costello:2018zrm} starts with a stack of $N$ D1-branes in $\mathbb{C}^3$. 
If we use coordinates $(x,y,z)$ on $\mathbb{C}^3$, they sit at $x=y=0$. We can take the probe D$1'$-brane 
to wrap the transverse line defined  $m_0 + x + u_0 y = 0$ and $z=z_0$ to engineer a determinant operator 
${\cal D}(m_0;u_0;z_0)$.\footnote{As a check, notice that if we displace the D1-branes to $x=X$ and $y=Y$, the combination
$m_0 + X + u_0 Y$ will have a zero eigenvalue whenever the D$1'$-brane intersects one of the D1-branes and the action of D$1-$D$1'$ strings correspondingly vanishes.} 

The backreaction of the D1-branes deforms the complex structure of the ambient manifold. The $x$, $y$ coordinates remain holomorphic. After a rescaling by a factor of $N^{\frac12}$ they will be identified with $b$ and $-a$ in $SL(2,\mathbb{C})$. The $z$ coordinate is not holomorphic anymore, but $(x z,y z)$ are deformed and rescaled by a factor of $N^{\frac12}$ to give holomorphic coordinates $d$ and $-c$ in $SL(2,\mathbb{C})$. 

The asymptotic shape of the probe D$1'$-brane for large $(a,b,c,d)$ is thus predicted to be
\begin{align}
\frac{b}{a} &=u_0 - \frac{m_i}{a} + \cdots\cr
\frac{c}{a} &=z_0 + \cdots \, .
\end{align}

We conclude that a D1-brane ${\cal B}_\rho$ supported on the spectral curve ${\cal S}_\rho$ has the correct shape at large $(a,b,c,d)$ to satisfy the holographic boundary condition for the insertion of the $k$ determinant operators ${\cal D}(m_1;u_1;z_1) \cdots {\cal D}(m_k;u_k;z_k)$. 

\subsection{The asymptotic phase space and the holographic action}
The world-volume theory of a single D1-brane is just a $\beta\gamma$ system where the two fields are valued in the 
normal bundle to the worldsheet. The combination $\beta \bar \partial \gamma$ is a $(0,1)$ form valued in the second exterior power of the normal bundle, which can be contracted with $\Omega$ to give a $(1,1)$ form which can be integrated over the curve to give the action of the $\beta\gamma$ system. 

Locally, we can parameterize the curve by the $a$ coordinate. We parameterize the normal fluctuations as fluctuations $\beta$ of $b$ and $\gamma$ of $c$. Using $\Omega = \frac{\dd a \,\dd b \,\dd c}{a}$ we get a standard action 
\begin{equation}
\int_{\cal S}  \beta \bar \partial \gamma \wedge \frac{\dd a}{a} .
\end{equation}
We find the classical phase space by expanding $\beta$ and $\gamma$ into powers of $a$:
\begin{align}
\beta(a) &= \sum_n \beta_n a^{n} \cr
\gamma(a) &= \sum_n \gamma_n a^{n} .
\end{align}
The Poisson brackets are
\begin{equation}
\{ \beta_n, \gamma_m \} = \delta_{n,-m} \qquad \qquad \{ \beta_n, \beta_m \} =\{ \gamma_n, \gamma_m \} =0 .
\end{equation}
As $a \to \infty$, the modes with positive $n$ are non-normalizable and are fixed by the holographic boundary conditions.
The mode $\beta_0$ is also fixed, as it represents a change of $m_i$. The conjugate mode $\gamma_0$ and 
the normalizable modes with negative $n$ are left free by the boundary conditions.

Given a solution of the classical equations of motion, the value of $\gamma_0$ will be the derivative of the classical action with respect to the fixed value of its conjugate variable $\beta_0$. Similarly, the modes with negative $n$ give the derivative of the classical action with respect to a change in the fixed values of the modes with positive $n$.\footnote{This is perfectly analogous to the statement that 
the classical mechanics action $S[x_i,x_f]$ evaluated on a solution with fixed initial and final positions $x_i$, $x_f$ determines the initial and final momenta as $p_f = \frac{\partial S}{\partial x_f}$ and $p_i = - \frac{\partial S}{\partial x_i}$.}

Applying this reasoning to each asymptotic end of the spectral curve, we learn that $\gamma_0=p_i$ at the $i$-th end represents the derivative of the brane action with respect to $m_i$. But $p_i$ also equals the derivative of the saddle action 
with respect to $m_i$. We thus get a full match at the leading order in $N$ between the action for a saddle $\rho$ of the (sub)determinant correlation function and the action of the holographic dual B-model brane ${\cal B}_\rho$, up to a function of $z_i$ and $u_i$ only. 

It is not too hard to match the $z_i$ and $u_i$ dependence of the actions as well, though it is a bit more cumbersome. 
For example, the $z_i$ derivative of the saddle action is 
\begin{equation}
\sum_{j\neq i}\frac{1}{u_i - u_j} \rho^i_j \rho^j_i .
\end{equation}
We need to compare this with the $a^{-1}$ term in the large $a$ expansion of $-b$. This is straightforward: the large $a$ expansion of $b$ is given by a systematic diagonalization of the matrix $-\mu + a^{-1} \rho$, which is just the non-degenerate perturbation theory in quantum mechanics. The $u_i$ play the role of unperturbed energies and the $\rho^i_j$ are the matrix elements of the perturbation. The expression above is the familiar second-order correction to the eigenvalues. 

We can match the $u_i$ derivatives in the same manner:
\begin{equation}
-\sum_{j\neq i}\frac{z_i-z_j}{(u_i - u_j)^2} \rho^i_j \rho^j_i = -\sum_{j\neq i}\frac{1}{z_i-z_j} [\rho^{-1}]^i_j [\rho^{-1}]^j_i
\end{equation}
matches with the $a^{-1}$ term in the large $a$ expansion of $c$ from the systematic diagonalization of $\zeta + a^{-1} \rho$.

\subsection{D-branes and spectral curves}
Observe that D-branes wrapping a spectral curve are a rather common occurrence in the B-model. A natural way to produce a complicated D-brane is to start with multiple copies of a simpler D-brane
and turn on matrix-valued vevs for the transverse deformations of the stack. Essentially, the world-volume theory on a stack of $k$ coincident D1-branes is a $\mathfrak{u}(k)$-gauged $\beta \gamma$ system valued in the adjoint representation. A classical solution gives commuting holomorphic vevs to $\beta$ and $\gamma$. 

Probing the system with another B-brane shows that the deformed stack of branes behaves as a single brane wrapping a spectral curve with transverse position given by the eigenvalues of $\beta$ and $\gamma$. The Chan-Paton bundle of the 
resulting D-brane is controlled by a natural line bundle on the spectral curve: the line bundle whose fiber consists of 
the common eigenvectors of $\beta$ and $\gamma$ with these eigenvalues. 

In the case at hand, it is natural to identify $B(a)$ and $C(a)$ with the matrix-valued vevs of fields $\beta(a)$ and $\gamma(a)$ 
describing the transverse positions of a stack of $k$ D1-branes which originally sit at $b=c=0$.\footnote{An alternative approach with equivalent results would be to describe a D1-brane as the result of tachyon condensation 
on a stack of space-filling D5-branes and anti-D5-branes, i.e. as a complex of sheaves $\mathbb{C}^k \to \mathbb{C}^{2k} \to \mathbb{C}^k$. One can build an appropriate differential out of commuting linear operators $b - B(a)$ and $c- C(a)$.} We will verify at the end of the section that the appropriate Chan-Paton bundle on ${\cal B}_\rho$ is indeed controlled by the eigenline bundle ${\cal L}_\rho$.

\subsection{Factorization and global symmetry}
The saddle solutions come automatically in $(\mathbb{C}^*)^k$ families, as we can conjugate $\rho$ 
by a diagonal invertible matrix to get a new solution. The spectral curve ${\cal S}_\rho$ is the same for all saddles in 
these continuous families. If we track back the origin of this symmetry, we see that it originates from a symmetry under rescaling of $\bar \psi_i$ and $\psi^i$ in opposite directions. We can interpret this as the remnant of the $U(1)$ gauge
symmetry on the bulk of the probe D$1'$ worldsheet, which becomes a $U(1)$ global symmetry at the boundary points
according to a standard holographic dictionary. 

We can consider saddle solutions such that $\rho$ is block-diagonal, with blocks $\rho_a$ of size $k_a$. Then the 
$\rho_a$ satisfy the saddle equations for the corresponding collection of $k_a$ determinant operators. 
The spectral curve decomposes correspondingly to the collection of ${\cal S}_{\rho_a}$, which are typically disjoint. 
This describes a configuration of disconnected D1-branes. Conversely, 
``connected'' holographic saddles associated to a single smooth semiclassical D1-brane correspond to ``irreducible'' 
solutions for which $\rho$ is not block diagonal. 

Recall that the chiral algebra correlation functions are covariant under both $SL(2)_L$ global conformal transformations and $SL(2)_R$ complexified R-symmetry transformations.  Our definition of the spectral curve, though, seems to treat the $a$ coordinate in a different way than the $b$, $c$ and $d$ coordinates. We will now show that the spectral curve transforms correctly under the action of $SL(2)_L\times SL(2)_R$ on $SL(2,\mathbb{C})$. 

The $Z(u;z)$ field transforms non-trivially under the action of $SL(2)_L\times SL(2)_R$: 
\begin{equation}
Z(u;z) = \frac{\gamma u + \delta}{\gamma' z + \delta'}Z\left( \frac{\alpha u + \beta}{\gamma u + \delta};   \frac{\alpha' z + \beta'}{\gamma' z + \delta'}\right) .
\end{equation}
We will thus also transform $m_i$ in the same manner so that the subdeterminants transform multiplicatively: 
\begin{equation}
{\cal D}(m;u;z) = \left(\frac{\gamma u + \delta}{\gamma' z + \delta'}\right)^N{\cal D}\left(\frac{\gamma' z + \delta'}{\gamma u + \delta}m; \frac{\alpha u + \beta}{\gamma u + \delta};   \frac{\alpha' z + \beta'}{\gamma' z + \delta'}\right) .
\end{equation}
Following through the derivation, the auxiliary $\rho$ variables transform as 
\begin{equation}
\rho \to \rho \frac{\gamma' \zeta + \delta'}{\gamma \mu + \delta} .
\end{equation}
Notice that there is a bit of latitude in the definition of the action of $SL(2)_L \times SL(2)_R$, as we can combine the above formula with conjugation by any diagonal matrix. 

Translations $\mu \to \mu + \beta$ or $\zeta \to \zeta + \beta'$ clearly act on the spectral curve as shifts $b \to b + \beta a$, $d \to d + \beta c$ or $c \to c + \beta' a$, $d \to d + \beta' b$. We thus only need to consider the action of  
inversions $\mu \to -\mu^{-1}$ or $\zeta \to - \zeta^{-1}$, as inversions and translations generate $SL(2)_L \times SL(2)_R$.
We can show the details for $\mu \to -\mu^{-1}$, as the analysis for $\zeta \to - \zeta^{-1}$ is essentially identical. 

Observe that an eigenvector $s$ of $B(a)$, etc. is annihilated by 
\begin{align}
&a \, \mu - \rho -b \cr
&a \, \zeta + \rho^{-1} -c\cr
&a \, \zeta \, \mu + \rho^{-1} \mu - \zeta \rho -d .
\end{align}
Then $\mu s$ is annihilated by 
\begin{align}
&b\, (-\mu^{-1}) - \rho \mu^{-1} + a \cr
&a \mu^{-1}\, \zeta + \rho^{-1} \mu^{-1} - c\mu^{-1}\cr
&a \, \zeta + \rho^{-1} - \zeta \rho \mu^{-1} -d \mu^{-1} ,
\end{align}
which is the same as being annihilated by 
\begin{align}
&b\, (-\mu^{-1}) - (\rho \mu^{-1})+ a \cr
&b \, \zeta  + (\rho \mu^{-1})^{-1}  -d \cr
&b\, \zeta  (-\mu^{-1}) +(\rho \mu^{-1})^{-1} (-\mu^{-1}) -  \zeta (\rho \mu^{-1})  + c .
\end{align}
We thus find that $-a$,$d$,$-c$ are simultaneous eigenvalues of 
\begin{align}
&b\, (-\mu^{-1}) - (\rho \mu^{-1}) \cr
&b \, \zeta  + (\rho \mu^{-1})^{-1} \cr
&b\, \zeta  (-\mu^{-1}) +(\rho \mu^{-1})^{-1} (-\mu^{-1}) -  \zeta (\rho \mu^{-1}),
\end{align}
i.e. the spectral curve transforms under inversion $\mu\rightarrow-\mu^{-1}$, $\rho \to \rho \mu^{-1}$ as: $a\rightarrow b, b\rightarrow -a, c\rightarrow d, d\rightarrow -c$ as desired.

\subsection{Off-diagonal fluctuations}
Consider now a reducible saddle solution with two blocks $\rho$ and $\rho'$. We may ask under which conditions on 
$\rho$ and $\rho'$ the reducible saddle admits infinitesimal block off-diagonal deformations.

Block off-diagonal fluctuations $\eta$ of the saddle solutions\footnote{We assume the saddle is of the form $\smqty( \rho & \eta \\ 0 & \rho')$.} satisfy a complicated-looking matrix equation:
\begin{equation} \label{eq:off}
\zeta \eta - \eta \zeta'  - \mu \rho^{-1} \eta (\rho')^{-1} + \rho^{-1} \eta (\rho')^{-1} \mu'=0 \, .
\end{equation}

We will now show that one can find a zeromode $\eta$ for every intersection point of ${\cal S}_\rho$ and ${\cal S}_{\rho'}$. The zeromode is built from the right eigenvector $s$ for $B(a)$, $C(a)$, $D(a)$ and the corresponding left eigenvector $s'$ for $B'(a)$, $C'(a)$, $D'(a)$. Let us take $\eta$ to be proportional to $\rho s \otimes s' +s \otimes s' \rho'$. Then, one can rewrite the left-hand side of \eqref{eq:off} in the following way\footnote{We thank Adri\'{a}n L\'{o}pez for pointing out a mistake here in the previous version.}
\begin{align}
    &(c\mu-d)s\otimes s'+\zeta s\otimes s'(a\mu'-b)-(a\mu-b)s\otimes s'-s\otimes s'(a\mu'-b)\zeta' \\
    -&\mu s\otimes s'(c-a\zeta')-\mu(c-a\zeta)s\otimes s'+s\otimes s'(d-b\zeta')+(c-a\zeta)s\otimes s'\mu' \, ,
\end{align}
where $b,c,d$ are the eigenvalues for $s$ as well as $s'$. After cancellations, we can arrange the remaining terms as
\begin{align}
    s\otimes s'\mqty[ b\zeta'+(c-a\zeta')\mu' ]-\mqty[ c\mu+\zeta(b-a\mu) ]s\otimes s' \, .
\end{align}
Both terms are equal to $d s\otimes s'$ so they cancel.

%This means that we can find a zeromode $\eta = \rho s \otimes s' +s \otimes s' \rho'$ for every intersection point of ${\cal S}_\rho$ and ${\cal S}_{\rho'}$. The zeromode is built from the corresponding right eigenvector $s$ for $B(a)$, $C(a)$, $D(a)$ and left eigenvector $s'$ for $B'(a)$, $C'(a)$, $D'(a)$.

We would like to employ this observation to compare the Chan-Paton bundle of the brane $B_\rho$ dual to some saddle $\rho$ 
with the canonical line bundle ${\cal L}_\rho$ present on any spectral curve, whose fiber is the common eigenline to $B(a)$, $C(a)$, $D(a)$. 

In order to ``see'' the Chan-Paton line bundle of a brane ${\cal B}_\rho$ we need to consider open strings stretched between 
${\cal B}_\rho$ and some other brane. We may employ another ${\cal B}_{\rho'}$ for this purpose, though 
B-model open strings will only appear if ${\cal B}_\rho$ and ${\cal B}_{\rho'}$ intersect at some point. 
As we deform ${\cal B}_{\rho'}$ and move the intersection point, the open strings will transform as a section of the 
Chan-Paton line bundle of ${\cal B}_\rho$, and viceversa. 

These open string modes are boundary local operators in the topological string worldsheet and generate the infinitesimal deformation of the two intersecting branes into a single smooth curve.
 It is thus natural to identify them with the solutions $\eta$ of (\ref{eq:off}) associated to the intersection point, which indeed transform as sections of ${\cal L}_\rho$ as we vary the intersection point.

This calculation also offers a subleading check of the holographic duality. In the presence of an off-diagonal zeromode, the 
semiclassical contribution of the block-diagonal saddle to the correlation function will diverge. This divergence arises from the exponentiation of an annulus diagram. On the holographic dual side, this diagram should correspond to the propagation of a closed string from ${\cal B}_\rho$ to ${\cal B}_{\rho'}$. It is easy to verify with a local calculation that this diagram in the B-model diverges in the correct manner when the two D-brane worldsheets intersect each other. 

\subsection{The inverse problem} \label{subsec:inverse}
Notice that our identification between semiclassical saddles goes only in one direction: we have built ${\cal B}_\rho$ from $\rho$, but we have not yet demonstrated that every B-model brane ${\cal B}$ which satisfies the correct boundary conditions can be produced from some $\rho$.\footnote{Considering the embedding of the chiral algebra calculation in the ${\cal N}=4$ SYM physical theory, it is not actually obvious that this should be the case: the physical brane configurations involve a supersymmetric D3-brane in AdS$_5 \times S^5$ and a B-model brane ${\cal B}$ in $SL(2,\mathbb{C})$ could potentially fail to lift to such a D3-brane. }

Often, spectral curve constructions can be inverted by a pushforward operation. Here we could try to foliate $SL(2,\mathbb{C})$ 
by surfaces of constant $a$ and pushforward the sheaf defining ${\cal B}$ along these surfaces. Concretely, that means 
that for generic $a$ we would consider the intersection points of the support  ${\cal S}$ of ${\cal B}$ with the constant $a$ surface 
and add up the fibers of the CP bundle ${\cal L}$ at these points. At branch points where two intersection points collide one needs to work a bit harder, but the basic idea is to produce a sheaf on the $a$ plane whose local sections in any open set $U$ are the same as the collection of local sections of ${\cal L}$ in the preimage of $U$ under the projection to the $a$ plane. 

Because of the boundary conditions on ${\cal B}$, the intersection points cannot move to infinity as we vary $a$. The number of intersection points will thus be generically fixed. Let us denote it by $k$. The direct sum of the fibers at intersection points 
will thus give a rank $k$ bundle. Commuting $k \times k$ matrices $B(a)$ and $C(a)$ could be defined as representing the multiplication action by $b$ and $c$ on the sections of ${\cal L}$.  A possible subtlety here is that the surfaces for generic $a$ are $\mathbb{C}^2$ parameterized by $b$ and $c$, but the fiber $\mathbb{C}\times \mathbb{C}^*$ at $a=0$ restricts $b c=1$ and leaves $d$ unconstrained. We expect that to simply impose the constraint $B(0) C(0)=0$.

We can trivialize the vector bundle at finite $a$ and write $B(a)$, $C(a)$ globally on the $a$ plane as polynomial matrices. 
We know that the eigenvalues of $B(a)$, $C(a)$ grow at infinity as $u_i a$ and $z_i a$. Unfortunately, 
this is not enough to immediately conclude that $B(a)$, $C(a)$ themselves should grow linearly at infinity. 
If it did, we would be done, as $B(a)$, $C(a)$ would have to take the form (\ref{eq:BCD}). \footnote{If we work with a single matrix $B(a)$, ignoring $C(a)$, we can easily produce counter-examples. The simplest one would be to write a matrix $B(a)$ with eigenvectors $(1,b,b^2, \dots, b^{k-1})$ for an eigenvalue $b$. 
Such a matrix would have all elements equal to $0$, $1$ or to the coefficients of the characteristic polynomial of $B(a)$. 
It would be of degree $k$ in $a$. In general, we can look for matrices such that the $(i,j)$ entry has degree up to $n_i-n_j+1$
for some collection of integers $n_i$. We can produce some examples of this sort, but we do not have a general understanding
of the situation. We leave it to future work. }

Instead, we will apply the inverse map to the special example of curves of genus $0$. Parameterize the curve by a global coordinate $t$
and denote as $t_i$ the location of the points which are mapped to infinity in $SL(2,\mathbb{C})$. We can parameterize
\begin{align}
a &= a_\infty + \sum_i \frac{a_i}{t-t_i} \cr
b &= b_\infty + \sum_i \frac{u_i a_i}{t-t_i} \cr
c &= c_\infty + \sum_i \frac{z_i a_i}{t-t_i} \cr
d &= d_\infty + \sum_i \frac{u_i z_i a_i}{t-t_i} \, .
\end{align}
We can compute 
\begin{equation}
a d - b c= a_\infty d_\infty- b_\infty c_\infty + \sum_i \frac{a_i}{t-t_i} \left[u_i z_i a_\infty- z_i b_\infty - u_i c_\infty+ d_\infty 
+ \sum_{j \neq i} a_j\frac{(u_i-u_j)(z_i - z_j)}{t_i - t_j} \right] \, .
\end{equation}
Therefore, we have constraints $a_\infty d_\infty- b_\infty c_\infty=1$ and 
\begin{equation}
u_i z_i a_\infty- z_i b_\infty - u_i c_\infty+ d_\infty + \sum_{j \neq i} a_j \frac{(u_i-u_j)(z_i - z_j)}{t_i - t_j} =0 \, .
\end{equation}
We can generically solve these $k$ linear equations for the $a_i$ in terms of $a_\infty$, $b_\infty$, $c_\infty$, $d_\infty$ and of the $t_i$.

Notice that we can act with fractional linear redefinitions of $t$. These will act on the $t_i$ in the obvious way, but will also shift $a_\infty$, etc. by multiples of the $a_i$. Using the solution for the $a_i$, one gets an intricate action on $a_\infty$, $b_\infty$, $c_\infty$, $d_\infty$.
At this point we thus have a $k$-dimensional space of curves which is the quotient of a $(k+3)$-dimensional space by $SL(2,\mathbb{C})$.
The point $(a_\infty, b_\infty, c_\infty, d_\infty)$ is the image of $t=\infty$, so under the reparameterization of $t$ the point $(a_\infty, b_\infty, c_\infty, d_\infty)$ will move along the curve. We  cannot use this $SL(2,\mathbb{C})$ to eliminate the $a_\infty$, $b_\infty$, $c_\infty$, $d_\infty$ parameters. We can use it instead to fix the values of $3$ of the $t_i$. 

We also find
\begin{align}
m_i &= u_i a_\infty-b_\infty + \sum_{j \neq i} \frac{u_i - u_j}{t_i-t_j} a_j \cr
p_i &= c_\infty - z_i a_\infty-  \sum_{j \neq i} \frac{z_i - z_j}{t_i-t_j} a_j .
\end{align}
Solving for the $a_\infty$, $b_\infty$, $c_\infty$, $d_\infty$ and $t_i$ as a function of $m_i$ seems hard. Instead, we can content ourselves with parameterizing the solutions by these parameters. 

Next, we will find an associated saddle $\rho$. We use the following trick. Consider the $k$-dimensional vector $s$ with components $\frac{1}{t-t_i}$ as a function on the curve.\footnote{This $s$ has a zero at $t=\infty$. The construction would work equally well if we rescale the components to 
$\frac{t-t_0}{t-t_i}$, placing the zero at some other point $t_0$. Even better, we should think about $s$ as a non-zero section of the spin bundle $K^{\frac12}$ on the genus $0$ curve.} We will now find a $\rho$ such that this is the common eigenvector of $B(a(t))$ and $C(a(t))$. 
Indeed, consider the vector $(a(t) \mu - b(t)) s$. As $a(t) u_i - b(t)$ is regular at $t=t_i$, the $i$-th entry of the vector has a simple pole at each $t_j$, including $t_i$. As $t \to \infty$, the vector goes to $0$. That means we can express uniquely each entry of $(a(t) \mu - b(t)) s$
as a linear combination of $\frac{1}{t-t_i}$. In other words, 
\begin{equation}
(a(t) \mu - b(t)) s = \rho s ,
\end{equation}
where $\rho^i_i = m_i$ and $\rho^i_j$ is computed as a residue, i.e.
\begin{equation}
\left[u_i a_\infty-b_\infty + \sum_{j \neq i} \frac{u_i - u_j}{t-t_j} a_j\right]\frac{1}{t-t_i}
\end{equation}
has residue at $t=t_j$: 
\begin{equation}
\rho^i_j = -\frac{u_i - u_j}{t_i-t_j} a_j .
\end{equation}
Similarly, we find $(\rho^{-1})^i_i = p_i$ and 
\begin{equation}
(\rho^{-1})^i_j = \frac{z_i - z_j}{t_i-t_j} a_j .
\end{equation}

We have thus found a natural ``genus $0$'' $k$-dimensional family of saddles $\rho$ parametrized by the $t_i$ and $a_\infty$, $b_\infty$, $c_\infty$, $d_\infty$ with $a_\infty d_\infty- b_\infty c_\infty=1$ modulo the $SL(2,\mathbb{C})$ action. 
It would be nice to know if this exhausts all saddles, or if higher genus saddles may exist. 

\section{Correlation functions of determinants and traces}\label{sec:corre}
In this section we consider correlation functions which include both a collection of (sub)determinant operators and a collection of single-trace operators. In the large $N$ expansion, these correlation functions will be controlled by the same large $N$ saddles as we found for the correlation function of subdeterminant operators: the single-trace insertions can at most produce some polynomial in $N$, which cannot compete with the exponentials associated to the determinants. 

We would like to show how the expansion around each large $N$ saddle takes the form of a standard 't Hooft expansion. 
We will begin by a small exercise in normal ordering which will facilitate the analysis. Consider the combination of all (sub)determinants which will appear in the correlation function, express it in a fermionic presentation and normal-order it with respect to the $Z$ contractions:
\begin{equation}
 \prod_i {\cal D}(m_i;u_i;z_i) =\int \prod_j [\dd \psi^j \dd \bar \psi_j] :e^{-\frac{1}{2N}\sum_{i \neq j} \frac{u_i-u_j}{z_i-z_j}\bar\psi_i \psi^j \bar\psi_j \psi^i +\sum_i \bar \psi_i \left(m_i+Z(u_i;z_i) \right) \psi^i}:_Z .
\end{equation}
This means that when we evaluate correlation function containing the expression on the right, we do not do Wick contractions between the 
$Z$'s in the exponent. 

We can now apply the Hubbard–Stratonovich transformation
\begin{equation}\label{eq:normalZ}
\prod_i {\cal D}(m_i;u_i;z_i) = Z_\rho^{-1} \int [\dd \psi \dd \bar\psi] [\dd \rho] :e^{\frac{N}{2}\sum_{i \neq j} \frac{z_i-z_j}{u_i-u_j}\rho^i_j \rho^j_i + \sum_{i,j} \rho^i_j \bar\psi_i \psi^j +\sum_i \bar \psi_i Z(u_i;z_i) \psi^i}:_Z .
\end{equation}
The normal order operation allowed us to introduce the auxiliary $\rho$ variables before integrating away the $Z$ fields. We can then
redefine $\rho \to \rho + \eta$ where $\eta$ is dynamical and $\rho$ is a background value which we can take to be any solution of the saddle equations (\ref{eq:saddle}), to insure the vanishing of the $\eta$ tadpole at the leading order in $N$.\footnote{At higher order in the $N^{-1}$ expansion it may be useful to adjust $\rho$ by subleading corrections.}

We are now ready for a diagrammatic analysis of the Feynman diagrams which arise from the $Z$, $\psi$, $\bar \psi$ and $\eta$ Wick contractions. We can use a standard ribbon graph picture: the $Z$ propagators are represented by ribbons with two colour lines,  
the fermion propagators by ribbons with one colour and one flavour line, the $\eta$ propagators by ribbons with two flavour lines.
The only constraint is that no $Z$ propagator can connect two $\bar \psi Z \psi$ vertices. Standard large $N$ 't Hooft combinatorics apply 
to these ribbon graphs, so that the power of $N$ is controlled by the topology of the surface which we obtain from the ribbon graph by filling in the closed colour lines: each connected component contributes $N^{2-2g-b}$, where $g$ is the genus and $b$ the number of closed flavour lines, which behave as boundaries of the surface. Determinant modifications can be readily included in the analysis. They behave as extra ``boundary'' vertices for the ribbon graphs.

Our main conjecture is that the large $N$ expansion of correlators around a saddle $\rho$ is holographically dual to a B-model calculation in the presence of a D-brane ${\cal B}_\rho$, order by order in the 't Hooft expansion. We will do various tests of this conjecture. 

\subsection{Determinants and a single-trace}
As our main example, consider the correlation function 
\begin{equation}
\langle {\cal D}(m_1;u_1;z_1) \cdots {\cal D}(m_k;u_k;z_k) N \Tr Z(u;z)^n \rangle .
\end{equation}
For simplicity, we assume that $z \neq z_i$ and $u \neq u_i$. 

For illustrative purposes, we will evaluate it in a standard way \citep{Jiang:2019xdz}, without the intermediate normal order step. We start from
\begin{equation}
\left\langle \prod_i {\cal D}(m_i;u_i;z_i) N \Tr Z(u;z)^n\right\rangle = \left\langle \int \prod_i [\dd \psi^i \dd \bar\psi_i]   e^{\bar\psi_i \left(m_i+Z(u_i;z_i) \right) \psi^i} N \Tr Z(u;z)^n\right\rangle,
\end{equation}
evaluate the $Z$ contractions
\begin{equation}
\int [\dd \psi \dd \bar\psi]\,  e^{-\frac{1}{2N}\sum_{i \neq j} \frac{u_i-u_j}{z_i-z_j}\bar\psi_i \psi^j \bar\psi_j \psi^i + \sum_i m_i \bar\psi_i \psi^i} N \Tr \left(-\frac{1}{N}\sum_i \psi^i \frac{u_i-u}{z_i-z} \bar\psi_i  \right)^n ,
\end{equation}
and reorganize the trace
\begin{equation}
- \int [\dd \psi \dd \bar\psi]\,  e^{-\frac{1}{2N}\sum_{i \neq j} \frac{u_i-u_j}{z_i-z_j}\bar\psi_i \psi^j \bar\psi_j \psi^i + \sum_i m_i \bar\psi_i \psi^i} N \Tr_{k \times k} \left(-\frac{1}{N}\bar\psi_j\psi^i \frac{u_i-u}{z_i-z}  \right)^n \, .
\end{equation}
Here the quantity in parenthesis is treated as a $k \times k$ matrix with indices $i$ and $j$.

We introduce the auxiliary $\rho$ variables\footnote{As before, we integrate over the $k(k-1)$ off-diagonal components of $\rho$ and $\rho^i_i=m_i$.}
\begin{equation}
- Z_\rho^{-1} \int [\dd \psi \dd \bar\psi] [\dd \rho]\, e^{\frac{N}{2}\sum_{i \neq j} \frac{z_i-z_j}{u_i-u_j}\rho^i_j \rho^j_i + \sum_{i,j} \rho^i_j \bar\psi_i \psi^j}N \Tr_{k \times k} \left(-\frac{1}{N}\bar\psi_j\psi^i \frac{u_i-u}{z_i-z}  \right)^n .
\end{equation}
As we execute the fermion Gaussian integral, the maximal power of $N$ will arise from contractions between each consecutive $\bar\psi_j\psi^i$ pair, giving an extra factor of $N$ for each. We thus obtain the leading $N$ answer
\begin{equation}
-  Z_\rho^{-1}\int [\dd \rho]\, e^{\frac{N}{2}\sum_{i \neq j} \frac{z_i-z_j}{u_i-u_j}\rho^i_j \rho^j_i } \left[\det \rho \right]^N N \Tr_{k \times k} \left(-\rho^{-1}\frac{\mu-u}{\zeta-z}  \right)^n .
\end{equation}
Evaluating this on some saddle $\rho$, the correlation function equals the ``bare'' correlation function of (sub)determinants 
$\langle {\cal D}(m_1;u_1;z_1) \cdots {\cal D}(m_k;u_k;z_k) \rangle$ times 
\begin{equation} \label{eq:onepoint}
- N \Tr_{k \times k} \left(-\rho^{-1}\frac{\mu-u}{\zeta-z}  \right)^n
\end{equation}
evaluated on the saddle. 

If we use the normal order trick, the diagrammatic interpretation of this answer is very straightforward. The leading ribbon graph is a disk diagram with the topology of a wheel, with the $n$ spokes being $Z$ propagators. Each $Z$ propagator gives one factor of $\frac{\mu-u}{\zeta-z}$ and each fermionic propagator gives a $-\rho^{-1}$ factor. The closed flavour line gives rise to the $\Tr_{k \times k}$.
The overall factor of $N$ appears because this is a disk diagram. 

\subsection{Geometric interpretation}
The disk one-point function (\ref{eq:onepoint}) can be computed in terms of the eigenvalues of the $k \times k$ matrix 
\begin{equation}
R(u;z) \equiv -\frac{1}{\zeta-z} \rho^{-1} (\mu-u) .
\end{equation}
We will now express this answer in the language of the spectral curve. We can write
\begin{equation}
R(u;z) = -\frac{1}{\zeta-z} \left(D(a) - u \, C(a) - \zeta B(a) + \zeta u a\right)
\end{equation}
for every $a$. Consider now the special values of $a$ where the spectral curve intersects the surface $\Delta(u;z)$ defined by $d - u c - z b + u z a =0$. There are $k$ such values: the solutions of the equation
 \begin{equation}
 \det \left[D(a) - u C(a) - z B(a) + u z a  \right] =0 \, .
 \end{equation}
The corresponding null eigenvectors are annihilated by $D(a) - u C(a) - z B(a) + u z a$. That means 
\begin{equation}
R(u;z) =  b - u a
\end{equation}
when acting on these eigenvectors. This means that the eigenvectors of $B(a)$, etc. at the intersection points 
are also eigenvectors of $R(u;z)$ with eigenvalues $b - u a$. 

As long as these $k$ eigenvectors are distinct, we can thus write 
\begin{equation}
\Tr_{k \times k} R(u;z)^n = \int_{S_\rho}  (b - u a)^n \, \delta_{\Delta(u;z)} \, ,
\end{equation}
where $\delta_{\Delta(u;z)}$ is a delta function supported on $\Delta(u;z)$. Indeed, the integral on the right localizes to the intersection points and gives the sum of the corresponding eigenvalues of $R(u;z)^n$. We expect this to happen for generic values of $u$ and $z$. 
By continuity, this relation must be true even when the eigenvectors are not distinct. 

\subsection{A holographic match}
Holographically, this disk 1-point function corresponds to a process where a closed string propagates from the boundary insertion corresponding to $N \Tr Z(u;z)^n$ to the $B_\rho$ brane dual to the saddle. This would evaluate to\footnote{See appendix F of \cite{Costello:2018zrm} for a review of the coupling of KS fields to D1-branes.}
\begin{equation}
\int_{S_\rho} \partial^{-1} \alpha_{n;u;z} ,
\end{equation}
where $ \alpha_{n;u;z}$ is the Kodaira-Spencer field sourced by the boundary insertion. 

This matches the chiral algebra calculation if we can identify 
\begin{equation}
\alpha_{n;u;z} = \partial \left[(b - u a)^n \delta_{\Delta(u;z)} \right] .
\end{equation}
This is indeed the case. The equation defining $\Delta(u;z)$ can also be written as 
\begin{equation}
 \frac{1}{a}\left(1+ (b - u a)(c - z a)\right)=0 .
\end{equation}
As we approach the boundary, the distribution is approximately supported on the two loci $\frac{c}{a}=z$ and $\frac{b}{a} = u$. 
We thus write 
\begin{equation}
\alpha_{n;u;z} \sim  \partial \left[(b - u a)^n \delta_{\frac{c}{a}=z} + (z a-c)^{-n} \delta_{\frac{b}{a}=u}\right] .
\end{equation}
The first term has a power law growth towards the boundary and is supported at the point $z$ of the holographic boundary. 
It has precisely the correct quantum numbers to represent the insertion of $N \Tr Z(u;z)^n$ at the boundary. 
In particular, it is a polynomial of degree $n$ in $u$, as expected.

The second term has a power law decay and represents the boundary behaviour of the field sourced by the $N \Tr Z(u;z)^n$ 
insertion. It would enter, say, in the calculation of a two-point function of $N \Tr Z(u;z)^n$ and another operator $N \Tr Z(u';z')^n$ at position $z' = \frac{c}{a}$.

We have thus matched the large $N$ expectation value of $A_n(u;z)$ in a saddle $\rho$ for a correlator of (sub)determinants 
to a B-model calculation of the same quantity in presence of the D1-brane ${\cal B}_\rho$. 

\subsection{From single-trace operators to modifications}
Next, consider the insertion of the global symmetry generator 
\begin{equation}
a_{n}(v;w) = \oint \frac{\dd z}{2 \pi i } \, (z-w)^{n-2} A_n(v;z) .
\end{equation}
As we vary $z$ along a closed loop, the codimension $2$ loci $\Delta(v;z)$ sweep a codimension $1$ locus $E(v)$, which intersects the spectral curve along some collection of loops. On $\Delta(v;z)$ we have $z = \frac{d-v c}{b - v a}$. At the leading order in $N$ in the saddle $\rho$, the insertion of $a_{n}(v;w)$ will thus produce a contour integral on these loops on ${\cal S}_\rho$:
\begin{equation}\label{eq:mod}
\oint \frac{1}{2 \pi i } \dd \frac{d-v c}{b - v a} \, \left(\frac{d-v c}{b - v a}-w\right)^{n-2} (b - v a)^n =n  \oint \frac{1}{2 \pi i } \omega_{J_n}(v;w).
\end{equation}

This is another remarkable test of the holographic correspondence. Indeed, the action of a symmetry generator in the B-model 
is implemented by cutting $SL(2,\mathbb{C})$ along a codimension $1$ locus $E$ and gluing it back under the action of 
the corresponding holomorphic vector field $V$. Equivalently, it corresponds to turning on a vev for the Kodaira-Spencer field $\alpha$ which is supported on $E$ proportional to $i_V \Omega$: 
\begin{equation}
\alpha = i_V \Omega \delta_E.
\end{equation} 
Recall that a D1-brane couples by $\partial^{-1} \alpha$. If $i_V \Omega = d \omega_V$, we can take 
\begin{equation}
\partial^{-1} \alpha = \omega_V \delta_E.
\end{equation} 
We have thus verified that the action of $a_{n}(v;w)$ on the chiral algebra side matches the action of $n J_n(v;w)$ on the B-model side:
the effect of the $J_n(v;w)$ vector fields on a D-brane supported on the spectral curve is precisely (\ref{eq:mod})!

Somewhat implicitly, this check also verifies the statement that the action of the global symmetry algebra onto a determinant operator 
produces modifications which are BRST-equivalent if the corresponding vector fields produce the same non-normalizable shape deformation 
in the dual D-brane. Indeed, it shows that the action of the symmetry generators is completely captured by the action of the corresponding vector fields onto the spectral curve. 
 
In the next section we will do a more direct test of this statement. 

\section{Two-point functions of modified determinants} \label{sec:test}
In this section we will focus on correlation functions in the presence of two determinants: $\det Y(\infty)$ and $\det X(0)$. 
We will employ the global symmetry generators to build a modification of both determinants and thus compute a matrix of
(large $N$) two-point functions of modified determinants. This will show explicitly which modifications are BRST-equivalent at leading order and match them to the action of vector fields on the B-model D1-brane. 

This correlation function has a single saddle, corresponding to the D-brane with support in the Cartan of $SL(2,\mathbb{C})$:\footnote{In appendix \ref{app:examples} we discuss saddles with a small number of determinants.}
\begin{equation}
g = \begin{pmatrix} a & 0 \cr 0 & \frac{1}{a}  \end{pmatrix} .
\end{equation}
A global symmetry generator $J_n(v;w)$, given by (\ref{eq:J}), deforms $g$ by
\begin{equation}
\delta g= \epsilon (a^{-1}+v w a)^{n-2} \begin{pmatrix} - v a & - v^2 a \cr \frac{1}{a} & \frac{v}{a}  \end{pmatrix} .
\end{equation}
We can combine this with the reparametrization $a \to a +\epsilon va(a^{-1}+v w a)^{n-2}$ to eliminate the variation of $a$ and $d$. We are left with variations 
\begin{equation}
\delta b(a) =- \epsilon v^2 a (a^{-1}+v w a)^{n-2}, \qquad \qquad \delta c(a) =\epsilon a^{-1} (a^{-1}+v w a)^{n-2} .
\end{equation}

The coefficient of $v^{k+1} w^{k-1}$ in $J_n(v;w)$ gives a variation 
\begin{equation}
\delta b(a) = - \epsilon {n-2 \choose k-1} a^{2k+1-n} ,
\end{equation}
which is a mode corresponding to a modification of $\det X(0)$ if $2k+1 \geq n$ and to a modification of $\det Y(\infty)$ otherwise. The coefficient of $v^k w^k$ in $J_n(v;w)$ gives a variation 
\begin{equation}
\delta c(a) = \epsilon {n-2 \choose k} a^{2k+1-n} ,
\end{equation}
which is a mode corresponding to a modification of $\det X(0)$ if $2k+1 > n$ and to a modification of $\det Y(\infty)$ otherwise.
All the other coefficients give vanishing variations. Notice that the modes $\delta b = a^s$ and $\delta c = a^{-s}$ for all $s$ are precisely dual under the inner product in the $\beta\gamma$ worldvolume theory of the D-brane.

We thus predict that the action of $a_n(v;w)$ on $\det X(0)$ will only produce very specific modifications at large $N$, up to BRST-exact 
operators:
\begin{itemize}
\item The coefficients of $v^{k+1} w^{k-1}$ for $2k+1 \geq n$ should give $-n {n-2 \choose k-1}$ times a modification 
which only depends on $2k+1-n$. 
\item The coefficients of $v^{k} w^{k}$ for $2k+1 > n$ should give $n {n-2 \choose k}$ times a modification 
which only depends on $2k+1-n$. 
\item All other coefficients should give BRST-exact or subleading modifications. 
\end{itemize}
The same statements apply to $\det Y(\infty)$, with the opposite constraints on $2k+1 - n$, producing modifications which are precisely dual under the inner product.

In the rest of the section we will often employ individual generators rather than generating functions of them. 
We can write
\begin{equation}
a^{(n)}_{p}(v) \equiv \sum_{q =-\frac{n}{2}}^{\frac{n}{2}} a^{(n)}_{p,q} v^{\frac{n}{2}-q}\equiv \oint \frac{\dd z}{2 \pi i z} \, z^{\frac{n}{2}+p} A_n(v;z) 
\end{equation}
i.e.
\begin{align}
a^{(n)}_{p,q} = \oint \frac{\dd z}{2\pi i z} z^{p+\frac{n}{2}} \oint\frac{\dd v}{2\pi i v} v^{q-\frac{n}{2}} A_n(v;z)
\end{align}
and then predict that the action of $a^{(n)}_{p,q}$ on $\det X(0)$ will only produce the following non-trivial modifications at large $N$: \begin{itemize}
\item $a^{(n)}_{p,p-1}$ for $2p-1 \leq 0$ should give $(-1)^{p-\frac{n}{2}}n$ times a modification 
which only depends on $p$. 
\item $a^{(n)}_{p,p+1}$ for $2p+1 < 0$ should give $(-1)^{p-1-\frac{n}{2}}n$ times a modification which only depends on $p$. 
\item All other coefficients should give BRST-exact modifications. 
\end{itemize}
The same statements apply to $\det Y(\infty)$, with the opposite constraints on $p$, producing modifications which are precisely dual under the inner product.

We will test these statements by computing correlation functions
\begin{equation} \label{eq:aadd}
\langle [a^{(m)}_{-p,-q}, \det Y(\infty)] [a^{(n)}_{p,q}, \det X(0)] \rangle
\end{equation}
in the large $N$ limit. We expect to find
\begin{equation} \label{eq:aa}
\frac{\langle [a^{(m)}_{-p,-q}, \det Y(\infty)] [a^{(n)}_{p,q}, \det X(0)] \rangle}{\langle \det Y(\infty) \det X(0) \rangle} \big|_{N\rightarrow\infty} = (-1)^{ \frac{m+n}{2}+1} n m N \left(\delta_{q,p-1} - \delta_{q,p+1}\right) .
\end{equation}
Notice that $n$ and $m$ are either both even or both odd in order for the answer to be non-zero. The relative sign in the parenthesis is due to the relative sign between inner products $\langle 0|\beta_n \gamma_{-n} |0 \rangle$
and $\langle 0|\gamma_n \beta_{-n} |0 \rangle$ in the $\beta\gamma$ system on the D1-brane.

\subsection{Chiral algebra computation}

The correlation functions (\ref{eq:aadd}) of modified determinants are computed by contour integrals from 
\begin{equation}
\langle \det Y(\infty) A_m(v';z') A_n(v;z) \det X(0)\rangle . \label{eq:DAAD}
\end{equation}
The general strategy to calculate correlation functions of two determinants and two single-trace operators is to employ the formula (\ref{eq:normalZ}), which in this case takes the form
\begin{align}
\det Y(\infty)\det X(0) = \frac{1}{Z_\rho} \int [\dd\psi\dd\bar\psi][\dd\rho] : e^{N\rho^1_2\rho^2_1 + \rho^1_2\bar\psi_1\psi_2 + \rho^2_1\bar\psi_2\psi_1 + \bar\psi_1 Y(\infty)\psi_1 + \bar\psi_2 X(0)\psi_2} :_Z .
\end{align}
We insert it to the correlation function\footnote{The normal ordered product $:\;:_Z$ introduced in section \ref{sec:corre}, means we do not contact symplectic bosons within the exponent.}
\begin{align}
& \langle \det Y(\infty) A_m(v';z') A_n(v;z) \det X(0) \rangle \\
& = \frac{1}{Z_\rho} \int [\dd\psi\dd\bar\psi][\dd\rho] e^{N\rho^1_2\rho^2_1} \expval{ : e^{\rho^1_2\bar\psi_1\psi_2 + \rho^2_1\bar\psi_2\psi_1 + \bar\psi_1 Y(\infty)\psi_1 + \bar\psi_2 X(0)\psi_2} :_Z  A_m(v';z') A_n(v;z) } .
\end{align}
We can study terms at the highest order in $N$ using diagrammatics explained in section \ref{sec:corre}. We need at least one contraction between each of the single-trace operators and the determinant operators in order for the contour integrals in (\ref{eq:aadd}) to not vanish. In the large $N$ limit, the dominant ribbon graphs are of order $N^1$ and have the topology of a disk with one fermion boundary. 

\input{diagram.tex}

All such possible diagrams are shown in figure \ref{fig:diag}. The two single-trace operators are connected by a certain number of ``inner'' $Z$ propagators:
\begin{align}
\wick{ \c1{Z}^i_j(v';z') \c1{Z}^k_l(v;z) = \delta^i_l\delta^k_j \frac{1}{N} \frac{v'-v}{z'-z} .}
\end{align}
The remaining fields are contracted with fermion vertices, $\bar\psi_1 Y(\infty)\psi_1$ or $\bar\psi_2 X(0)\psi_2$, along the unique closed flavour loop. As we go along the flavour loop, we encounter first the sequence of fermionic vertices contracted with the first single-trace operator $A_m(v';z')$ and then the sequence of fermionic vertices contracted with the second single-trace operator $A_n(v;z)$. The contractions are:
\begin{align}
\wick{ \c1{X}^i_j(0) \c1{Z}^k_l(v;z) = \delta^i_l\delta^k_j \frac{1}{N} \frac{v}{z} , \qquad \c2{Y}^i_j(\infty) \c2{Z}^k_l(v;z) = \delta^i_l\delta^k_j \frac{1}{N} .}
\end{align}

The $A_n(v;z)$ operator can be ``inserted'' in such a diagram in $n$ different ways, depending on which of the $Z$ fields is connected to the ``first'' propagator in the diagram. The same is true for $A_m(v';z')$. The resulting $nm$ combinatorial factor already appears in the answer (\ref{eq:aa}). The sum over diagrams will have to reproduce the remaining $\delta_{q,p-1} - \delta_{q,p+1}$ and appropriate overall sign. 

Multiple diagrams with the same topology are distinguished by two things:
\begin{itemize}
\item The number of inner $Z$ propagators between single-trace operators $A_m(v';z')$ and $A_n(v;z)$, which we denote $\alpha$.
\item Which of the two flavours of fermions run in the two propagators in between the two sequences of fermionic vertices.
\end{itemize}

The fermion propagator is completely off-diagonal, so the two flavours of fermions have to alternate along the boundary. Let us denote by $\varepsilon$ the difference between the numbers of contractions $\wick{\c{X}(0)\c{Z}(v';z')}$ and $\wick{\c{Y}(\infty)\c{Z}(v';z')}$.  When $m-\alpha$ and $n-\alpha$ are both even there are two possible diagrams, both with $\varepsilon=0$. They give equal contributions. When $m-\alpha$ and $n-\alpha$ are odd there are two distinct diagrams, with $\varepsilon=1$ and $\varepsilon=-1$.

A diagram characterized by $(\alpha,\varepsilon)$ gets the contribution from all propagators:
\begin{align}
(-1)^{\frac{m+n}{2}-\alpha+1}\qty(\frac{v'-v}{z'-z})^\alpha \qty(\frac{v'}{z'})^{\frac{m-\alpha-\varepsilon}{2}} \qty(\frac{v}{z})^{\frac{n-\alpha+\varepsilon}{2}} ,
\end{align}
where the sign comes from fermion contractions. When inserted into the contour integrals it produces two binomial factors:
\begin{align}
\oint_\infty\frac{\dd z'}{2\pi i z'} z'^{\frac{m}{2}-p} & \oint_\infty\frac{\dd v'}{2\pi i v'} v'^{-q-\frac{m}{2}} \oint_0\frac{\dd z}{2\pi i z} z^{\frac{n}{2}+p} \oint_0\frac{\dd v}{2\pi i v} v^{q-\frac{n}{2}} \, \qty(\frac{v'-v}{z'-z})^\alpha \qty(\frac{v'}{z'})^{\frac{m-\alpha-\varepsilon}{2}} \qty(\frac{v}{z})^{\frac{n-\alpha+\varepsilon}{2}} \\
&= (-1)^{\frac{\alpha-\varepsilon}{2}-q} \binom{\alpha}{\frac{\alpha+\varepsilon}{2}+q} \binom{\frac{\alpha+\varepsilon}{2}-p-1}{\alpha-1} \equiv s(\alpha,\varepsilon) . \label{eq:s}
\end{align}

The final answer is the sum of three contributions, each of which is a sum over the possible values of $\alpha$. Schematically, 
\begin{align} 
 (-1)^{\frac{m+n}{2}+1}\qty( 2\!\!\!\!\sum_{m-\alpha\text{ even}} (-1)^\alpha s(\alpha,0) + \sum_{m-\alpha\text{ odd}} (-1)^\alpha s(\alpha,-1) + \sum_{m-\alpha\text{ odd}} (-1)^\alpha s(\alpha,1) ) . \label{eq:sum3}
\end{align}
It turns out that the combination in parenthesis is non-vanishing only in two cases:\footnote{The computation is relegated to the appendix \ref{app:B}. There is actually a third non-vanishing case, with $p=q=0$: the $a^{(n)}_{0,0}$ generators do not 
create a modification of the determinant, but they give back a multiple of the determinant itself.}
\begin{align}
=\begin{cases}
    +1 & p-q=1, p\leq \frac{1}{2} \\
    -1 & q-p=1, p< -\frac{1}{2} .
\end{cases}
\end{align}
We arrive at the expected
\begin{equation} \label{eq:aa2}
\frac{\langle [a^{(m)}_{-p,-q}, \det Y(\infty)] [a^{(n)}_{p,q}, \det X(0)] \rangle}{\langle \det Y(\infty) \det X(0) \rangle} \big|_{N\rightarrow\infty} = (-1)^{\frac{m+n}{2}+1} n m N \left(\delta_{q,p-1} - \delta_{q,p+1}\right) .
\end{equation}

Since the modifications by $a^{(n)}_{p,q}$ do not depend on $n$ up to a number, we can compute them using the lowest possible $n$: 
\begin{align*}
[a^{(-2p+2)}_{p,p-1}, \det X(0)] &= \oint_0\frac{\dd z}{2\pi i} \oint_0\frac{\dd v}{2\pi i v} v^{2p-2} N\Tr Z(v;z)^{-2p+2} \det X(0) 
\end{align*}
which gives a substitution $X \to (2-2p) Y^{1-2p}$. The second type of modification is\begin{align}
[a^{(-2p+2)}_{p,p+1}, \det X(0)] &= \oint_0\frac{\dd z}{2\pi i} \oint_0\frac{\dd v}{2\pi i v} v^{2p} N\Tr Z(v;z)^{-2p+2} \det X(0) .
\end{align}
There are two terms which contribute at large $N$, schematically:
\begin{align}
\varepsilon\varepsilon(X,\dots,X,Y^{-2p-2}\partial X) + \varepsilon\varepsilon(X,\dots,X,\partial^2 Y^{-2p-3}) .
\end{align}

\section{The matrix model configuration}\label{sec:matrix}
This section is somewhat orthogonal to the rest of the paper. It describes a topological subsector where the chiral algebra correlation functions
reduce to correlation functions in a Gaussian matrix model. This subsector may be related the Dijkgraaf-Vafa setup \cite{Dijkgraaf:2002dh,Dijkgraaf:2002fc,Aganagic:2003qj}.

The definition of the topological subsector is akin to the manner the chiral algebra itself is embedded into the physical theory.
Recall that the chiral algebra is embedded in the physical theory by realizing $X$ and $Y$ as position-dependent 
linear combinations of scalar fields of the physical theory. Concretely, one has four scalars $\phi_{\alpha \dot \alpha}$ 
and takes combinations $X = \phi_{1 \dot 1} + \bar z \phi_{1 \dot 2}$ and $Y = \phi_{2 \dot 1} + \bar z \phi_{2 \dot 2}$.

We can imitate this construction to define a subsector of the chiral algebra generated by a field 
\begin{equation}
M(z) \equiv X(z) + z Y(z) = Z(u=z;z) .
\end{equation}
Correlation functions of $M(z_i)$ are independent of the positions $z_i$. Wick contractions of $M$ with itself 
are identical to these of a Gaussian matrix model. 

One way to understand the origin of this simplification is that the correlation functions of operators built from $M(z)$ preserve some extra superconformal symmetry, i.e. the fermionic linear combinations of symmetry generators:
\begin{equation}
\oint \frac{\dd z}{2 \pi i} \Tr b Z(z;z) 
\end{equation}
and 
\begin{equation}
\oint \frac{\dd z}{2 \pi i} \Tr \partial c Z(z;z) \, .
\end{equation}

In general, we can use the $SL(2)_R$ symmetry to rotate at least three $u_i$ to coincide with some $z_i$. As a result, 
two- and three-point functions can always be taken to lie in the protected subsector. This can be computationally useful. 

\subsection{Protected determinants}

As soon as we set $z_i = u_i$, the saddle equations reduce to 
\begin{equation}
\rho + \rho^{-1} = \lambda
\end{equation}
for some diagonal matrix $\lambda$. For irreducible solutions, $\lambda$ has to be a multiple of the identity. 

In an irreducible solution, $\rho$ will have $k_+$ eigenvalues equal to some $r$ and $k_-$ eigenvalues equal to $r^{-1}$,
with $r + r^{-1} = \lambda$. Before imposing the $m_i = \rho_i^i$ constraint, we have a $2 k_+ k_-$-dimensional space of 
solutions. Imposing $m_i = \rho_i^i$ enforces $k_+ r + k_- r^{-1} = \Tr m$ and removes $k-1$ degrees of freedom. 
If we also quotient by the identification $\rho \to \lambda \rho \lambda^{-1}$ for diagonal $\lambda$, we get 
a $2(k_+-1)(k_--1)$-dimensional space of solutions. For the $k=1,2,3$ cases 
the space has expected dimension $0$.

As we adjust the spectral problem to the simplified setting, we find 
\begin{align}
B(a) &= a \, \zeta - \rho \cr
C(a) &= a \, \zeta - \rho + \lambda \cr
D(a) &= a \, \zeta^2 - \rho \zeta- \zeta \rho + \lambda \zeta
\end{align}
and thus an irreducible spectral curve is supported on $c-b = \lambda$.

As we compute correlation functions of operators with $u_i = z_i$, the Wick contractions between different operators 
are the same as in a Gaussian matrix model. We still do not have contractions between scalar fields in the same operator, 
which means that the chiral algebra correlation functions reduce to matrix model correlation functions of {\it normal ordered}
operators.

There is a simple integral relation between a normal-ordered determinant and a standard determinant in the matrix model. 
If we start from the fermionized form 
\begin{equation}
:\det \left( \lambda + M \right): =\int \dd \psi  \dd \bar \psi :e^{\bar \psi \left(\lambda +M \right) \psi}:
\end{equation} 
then we can remove the normal ordering by writing 
\begin{equation}
\det \left( \lambda + M \right) =\int \dd \psi  \dd \bar \psi :e^{\bar \psi \left(\lambda +M \right) \psi}:e^{-\frac{1}{2N} (\bar \psi\psi)^2}
\end{equation} 
and then
\begin{equation}
\det \left( \lambda + M \right) =\sqrt{-\frac{N}{2 \pi}} \int \dd \psi  \dd \bar \psi \dd m :e^{\bar \psi \left(m + M \right) \psi}:e^{\frac{N}{2} (m - \lambda)^2} .
\end{equation} 
At the level of saddle equations, this means that instead of fixing $m_i$ we fix $m_i + p_i = \lambda_i$. 

The saddle equations are unchanged, but now the diagonal components of $\rho$ are free and $\lambda$ is fixed. 
We only get irreducible solutions if all the $\lambda_i$ happen to coincide. Otherwise, the only saddles are completely diagonal 
and the $k$ branes $S_{m_i}$ are disconnected from each other.  

%Notice that $c-b$ is invariant under the diagonal $SL(2)_d \subset SL(2)_L \times SL(2)_R$ symmetry which preserves the 
%$u_i = z_i$ constraint. Indeed, the combination $c-b$ is invariant under an infinite subalgebra of the global symmetry algebra:
%\begin{equation}
%J_2(v)(c-b) = d - v c - v b + v^2 a
%\end{equation}
%and 
%\begin{equation}
%I_0(w)(c-b) = - (d - w b - w c + w^2 a)
%\end{equation}
%imply that $J_n(v;v)+ I_n(v;v)$ fix $c-b$. 

\subsection*{Acknowledgements}

We thank Kevin Costello and Ben Webster for useful discussions. This research was supported in part by a grant from the Krembil Foundation. K.B. and D.G. are supported by the NSERC Discovery Grant program and by the Perimeter Institute for Theoretical Physics. Research at Perimeter Institute is supported in part by the Government of Canada through the Department of Innovation, Science and Economic Development Canada and by the Province of Ontario through the Ministry of Colleges and Universities. 

\appendix
\section{Examples} \label{app:examples}
\subsection{$k=1$}
The saddle equations are trivial. We have $\rho = m_1$ and thus $p_1 = m_1^{-1}$. This is reasonable: a one-point function 
of ${\cal D}(m_1;u_1;z_1)$ can only pick the $m_1^N$ term, so the saddle action is $\log m_1$. 

The spectral curve is given by $b=u_1 a - m_1$, $c = z_1 a + m_1^{-1}$, $d = u_1 z_1 a + u_1  m_1^{-1} - z_1 m_1$,
i.e. 
\begin{equation}
g = \begin{pmatrix} 1 & u_1 \cr z_1 & u_1 z_1 \end{pmatrix} a +  \begin{pmatrix} 0 & -m_1 \cr m_1^{-1} & (u_1 m_1^{-1} - z_1 m_1) \end{pmatrix} .
\end{equation}

If we identify $t=a$, this is a standard genus 0 spectral curve with one puncture at $t=\infty$.
\subsection{$k=2$}
For an irreducible solution, the saddle equations require
\begin{align}
-\rho^1_2 \rho^2_1 = \frac{u_1-u_2}{z_1-z_2}-m_1m_2
\end{align}
and we have 
\begin{equation}
p_1 =- \frac{z_1-z_2}{u_1-u_2} m_2, \qquad \qquad p_2 =- \frac{z_1-z_2}{u_1-u_2} m_1.
\end{equation}
The two-point function $\langle {\cal D}(m_1;u_1;z_1) {\cal D}(m_2;u_2;z_2)\rangle$ has a tree-level saddle action 
\begin{equation}
\log \frac{z_1-z_2}{u_1-u_2}- 1-m_1m_2\frac{z_1-z_2}{u_1-u_2}.
\end{equation}

The spectral curve is of genus $0$. We can place the punctures at $t=0$ and $t=\infty$ and write without loss of generality:
\begin{equation}
g = \begin{pmatrix} 1 & u_1 \cr z_1 & u_1 z_1 \end{pmatrix} t + \begin{pmatrix} 1 & u_2 \cr z_2 & u_2 z_2 \end{pmatrix} \frac{a_2}{t}+   \begin{pmatrix} a_\infty & b_\infty \cr c_\infty & d_\infty \end{pmatrix}.
\end{equation}
The constraint $\det g=1$ fixes all but two parameters. 

As we placed a puncture at $t=\infty$,  we should replace the basis $\frac{1}{t-t_i}$ of functions vanishing at $t=\infty$ which we employed in the general case in section \ref{subsec:inverse}
with some functions that vanish at some other point $t_0$ instead of $t=\infty$. If we use $(t-t_0)$ and $(t-t_0)/t$ as entries of the eigenvector, we arrive 
to the expected form of $\rho$, with $\rho^2_1 = u_1 - u_2$, and can determine the two remaining parameters in terms of $m_1$ and $m_2$. 
We get
\begin{align}
a_\infty &=- \frac{m_1-m_2}{u_1-u_2} \cr
b_\infty &=- \frac{m_1 u_2-m_2 u_1}{u_1-u_2} \cr
c_\infty &=- \frac{m_1 z_1-m_2 z_2}{u_1-u_2} \cr
a_\infty &=- \frac{m_1 u_2 z_1-m_2 u_1 z_2}{u_1-u_2} \cr
a_2 &= -\frac{\rho^1_2 \rho^2_1}{(u_1-u_2)^2} .
\end{align}

\subsection{$k=3$}
The $k=3$ saddle equations have generically two irreducible solutions. The solutions involve a square root and are cumbersome to write down directly. 
Instead, we can realize them in terms of genus $0$ spectral curves. 
We can write
\begin{equation}
g = \begin{pmatrix} 1 & u_1 \cr z_1 & u_1 z_1 \end{pmatrix} a_1 t + \begin{pmatrix} 1 & u_2 \cr z_2 & u_2 z_2 \end{pmatrix} \frac{a_2}{t}+ \begin{pmatrix} 1 & u_3 \cr z_3 & u_3 z_3 \end{pmatrix} \frac{a_3}{t-1}+   \begin{pmatrix} a_\infty & b_\infty \cr c_\infty & d_\infty  \end{pmatrix}.
\end{equation}
The constraint $\det g=1$ allows one to express, say, $a_2$, $a_3$, $c_\infty$ and $d_\infty$ in terms of $a_1$, $a_\infty$, $d_\infty$.

We can build $\rho$ by using the eigenvector with components $(t-t_0)$, $(t-t_0)/t$, $(t-t_0)/(t-1)$. This expresses the masses $m_i$ in terms of 
$a_1$, $a_\infty$, $d_\infty$. The relations are linear in $a_\infty$, $d_\infty$, but impose a quadratic equation for $a_1$. 
We thus find a nice parameterization of the saddles $\rho$ in terms of $a_1$ and two of the $m_i$, together with a quadratic relation to 
impose the value of the third $m_i$.

\section{Binomial sum}
\label{app:B}

After change of variables $x=\frac{\alpha+\varepsilon}{2}+q$ in (\ref{eq:s}), the three sums in (\ref{eq:sum3}) take the form:
\begin{align}
 S(\varepsilon) = \sum_{x=\lceil \max( q+\varepsilon/2,2q+\varepsilon)\rceil}^{q-p+\varepsilon} (-1)^x \binom{x-q-p-1}{2x-2q-\varepsilon-1} \binom{2x-2q-\varepsilon}{x} ,
\end{align}
where $\varepsilon\in\{0,\pm 1\}$ and the total sum is equal
\begin{align}
(-1)^{\frac{m+n}{2}+1}(2S(0) + S(-1) + S(1) ).
\end{align}

We will use the binomial identity:
\begin{align}
    S(B,C) &\equiv \sum_{x=0}^{B} (-1)^{x} \binom{x+C-1}{2x-B+C-1} \binom{2x-B+C}{x} = \begin{cases} 1, & B=0 \\ 2(-1)^B, & B>0 . \end{cases} \label{eq:id}
\end{align}

There are 9 ranges for variables $p-q$ and $p+q$ that we have to consider:
\begin{enumerate}
\item $p-q\geq 2$: all three sums are zero.
\item $p-q=1,p+q\leq 0$:
% only $S(1)$ is not empty and by (\ref{eq:id}) it equals 1:
\begin{align}
S(0)=0, \qquad S(-1)=0, \qquad S(1)=1 .
\end{align}
\item $p-q=0,p+q\leq -1$: 
%$S(-1)$ is empty and by (\ref{eq:id}):
\begin{align}
\qquad S(0) = 1, \qquad S(-1)=0, \qquad S(1) = -2 .
\end{align}
\item $p-q=0,p+q=0$:
\begin{align}
S(0)=1, \qquad S(-1)=0, \qquad S(1)=-1 .
\end{align}
\item $p-q=-1,p+q\leq -1$: 
%by (\ref{eq:id}):
\begin{align}
S(0) = -2, \qquad S(-1) = 1, \qquad S(1) = 2 .
\end{align}
\item $p-q=-1,p+q=0$:
\begin{align}
S(0)=-1, \qquad S(-1)=1, \qquad S(1)=1 .
\end{align}
\item $p-q\leq -2,p+q\leq -1$: 
%by (\ref{eq:id}):
\begin{align}
S(0) = 2(-1)^{q-p}, \qquad S(-1) = -2(-1)^{q-p}, \qquad S(1) = -2(-1)^{q-p} .
\end{align}
\item $p-q\leq -2,p+q =0$: 
%the sums start from $x=1$ so we subtract from $2(-1)^{q-p+\varepsilon}$ the element of the sum with $x=0$:
\begin{align}
S(0) = (-1)^{2q}, \qquad S(-1) = -(-1)^{2q}, \qquad S(1) = -(-1)^{2q}  .
\end{align}
\item $p+q\geq 1$: all sums are zero again.
\end{enumerate}

The total sum is non-zero only for:
\begin{align}
2S(0)+S(-1)+S(1) = \begin{cases} +1, & p-q=1,p\leq \frac{1}{2} \\ -1, & p-q=-1,p < \frac{1}{2} \\ +1, & p=q=0, \end{cases}
\end{align}
where the last case does not correspond to a modification of a determinant.

\bibliographystyle{JHEP}

\bibliography{mono}

\end{document}

%% file: GGpic.tex
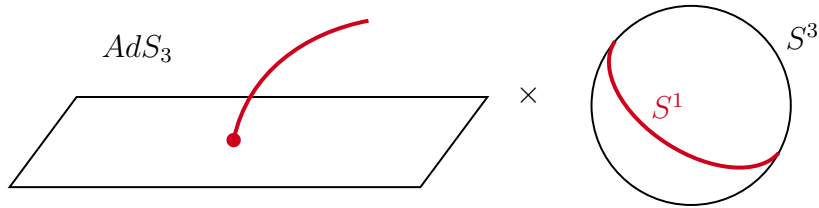
\begin{figure}
\centering

\tikzset{every picture/.style={line width=0.75pt}} %set default line width to 0.75pt        

\begin{tikzpicture}[x=0.75pt,y=0.75pt,yscale=-1,xscale=1]
%uncomment if require: \path (0,655); %set diagram left start at 0, and has height of 655

%Shape: Parallelogram [id:dp8005719421525677] 
\draw   (138.8,283) -- (346,283) -- (312.2,328.5) -- (105,328.5) -- cycle ;
%Curve Lines [id:da9457419442888824] 
\draw [color={rgb, 255:red, 208; green, 2; blue, 27 }  ,draw opacity=1 ][line width=1.5]    (218,304.75) .. controls (225,270.75) and (254,250.5) .. (286,244.5) ;
\draw [shift={(218,304.75)}, rotate = 281.63] [color={rgb, 255:red, 208; green, 2; blue, 27 }  ,draw opacity=1 ][fill={rgb, 255:red, 208; green, 2; blue, 27 }  ,fill opacity=1 ][line width=1.5]      (0, 0) circle [x radius= 2.61, y radius= 2.61]   ;
%Shape: Circle [id:dp6115745876241541] 
\draw   (398.5,287.25) .. controls (398.5,259.5) and (421,237) .. (448.75,237) .. controls (476.5,237) and (499,259.5) .. (499,287.25) .. controls (499,315) and (476.5,337.5) .. (448.75,337.5) .. controls (421,337.5) and (398.5,315) .. (398.5,287.25) -- cycle ;
%Shape: Arc [id:dp184036369761291] 
\draw  [draw opacity=0][line width=1.5]  (492.89,311.21) .. controls (483.18,322.61) and (458.71,320.76) .. (436.58,306.52) .. controls (413.34,291.57) and (401.47,268.61) .. (410.07,255.25) .. controls (410.16,255.12) and (410.24,254.99) .. (410.32,254.87) -- (452.15,282.32) -- cycle ; \draw  [color={rgb, 255:red, 208; green, 2; blue, 27 }  ,draw opacity=1 ][line width=1.5]  (492.89,311.21) .. controls (483.18,322.61) and (458.71,320.76) .. (436.58,306.52) .. controls (413.34,291.57) and (401.47,268.61) .. (410.07,255.25) .. controls (410.16,255.12) and (410.24,254.99) .. (410.32,254.87) ;

% Text Node
\draw (359,275.4) node [anchor=north west][inner sep=0.75pt]    {$\times $};
% Text Node
\draw (150,250.4) node [anchor=north west][inner sep=0.75pt]    {$\mathrm{AdS}_{3}$};
% Text Node
\draw (495,244.4) node [anchor=north west][inner sep=0.75pt]    {$S^{3}$};
% Text Node
\draw (427,279.4) node [anchor=north west][inner sep=0.75pt]  [color={rgb, 255:red, 208; green, 2; blue, 27 }  ,opacity=1 ]  {$S^{1}$};

\end{tikzpicture}

\caption{A schematic depiction of a Giant Graviton $D$1-brane wrapping $\mathbb{C}^*$ in Euclidean AdS$_3\times S^3 \approx SL(2,\mathbb{C}).$}
\label{fig:GG}
\end{figure}

%% file: diagram.tex
\begin{figure}[t]
\centering
\begin{tikzpicture}

%\draw[scale=1.1,thick,dashed] (0,0) ellipse (4.5cm and 3.5cm);
%\draw (0,0) ellipse (4.5cm and 3.5cm);

%UP
\coordinate (a1) at ({-2+cos(10)} , {sin(10)} );
\coordinate (a2) at ({-2+cos(30)} , {sin(30)} );
\coordinate (a3) at ({-2+cos(50)} , {sin(50)} );
\coordinate (a4) at ({-2+cos(70)} , {sin(70)} );

\coordinate (a5) at ({-2+cos(90)} , {sin(90)} );
\coordinate (a6) at ({-2+cos(110)} , {sin(110)} );
\coordinate (a7) at ({-2+cos(130)} , {sin(130)} );
\coordinate (a8) at ({-2+cos(150)} , {sin(150)} );

\draw[thick] (a2) arc (30:50:1cm);
\draw[thick] (a4) arc (70:90:1cm);
\draw[thick] (a6) arc (110:130:1cm);
\draw[thick] (a8) arc (150:170:1cm);

%DOWN
\coordinate (ad1) at ({-2+cos(-10)} , {sin(-10)} );
\coordinate (ad2) at ({-2+cos(-30)} , {sin(-30)} );
\coordinate (ad3) at ({-2+cos(-50)} , {sin(-50)} );
\coordinate (ad4) at ({-2+cos(-70)} , {sin(-70)} );

\coordinate (ad5) at ({-2+cos(-90)} , {sin(-90)} );
\coordinate (ad6) at ({-2+cos(-110)} , {sin(-110)} );
\coordinate (ad7) at ({-2+cos(-130)} , {sin(-130)} );
\coordinate (ad8) at ({-2+cos(-150)} , {sin(-150)} );

\draw[thick] (ad2) arc (330:310:1cm);
\draw[thick] (ad4) arc (290:270:1cm);
\draw[thick] (ad6) arc (250:230:1cm);
\draw[thick] (ad8) arc (210:190:1cm);

%second circle
%UP
\coordinate (b1) at ({2+cos(170)} , {sin(170)} );
\coordinate (b2) at ({2+cos(150)} , {sin(150)} );
\coordinate (b3) at ({2+cos(130)} , {sin(130)} );
\coordinate (b4) at ({2+cos(110)} , {sin(110)} );

\coordinate (b5) at ({2+cos(90)} , {sin(90)} );
\coordinate (b6) at ({2+cos(70)} , {sin(70)} );
\coordinate (b7) at ({2+cos(50)} , {sin(50)} );
\coordinate (b8) at ({2+cos(30)} , {sin(30)} );

\draw[thick] (b2) arc (150:130:1cm);
\draw[thick] (b4) arc (110:90:1cm);
\draw[thick] (b6) arc (70:50:1cm);
\draw[thick] (b8) arc (30:10:1cm);

%DOWN
\coordinate (bd1) at ({2+cos(-170)} , {sin(-170)} );
\coordinate (bd2) at ({2+cos(-150)} , {sin(-150)} );
\coordinate (bd3) at ({2+cos(-130)} , {sin(-130)} );
\coordinate (bd4) at ({2+cos(-110)} , {sin(-110)} );

\coordinate (bd5) at ({2+cos(-90)} , {sin(-90)} );
\coordinate (bd6) at ({2+cos(-70)} , {sin(-70)} );
\coordinate (bd7) at ({2+cos(-50)} , {sin(-50)} );
\coordinate (bd8) at ({2+cos(-30)} , {sin(-30)} );

\draw[thick] (bd2) arc (210:230:1cm);
\draw[thick] (bd4) arc (250:270:1cm);
\draw[thick] (bd6) arc (290:310:1cm);
\draw[thick] (bd8) arc (330:350:1cm);

%elipse nodes
\coordinate (e1) at ($(0,0)+(130:4.5 and 3.5)$);
\coordinate (e2) at ($(0,0)+(135:4.5 and 3.5)$);
\draw[thick] (e2) arc
    [
        start angle=135,
        end angle=151,
        x radius=4.5cm,
        y radius =3.5cm
    ] ;
\coordinate (h1) at ($(0,0)+(143:4.5*1.1 and 3.5*1.1)$);
\filldraw[black] (h1) circle (1.5pt);
\coordinate (e3) at ($(0,0)+(151:4.5 and 3.5)$);
%\draw[dashed, thick] (e3) arc
%    [
%        end angle=143,
 %       x radius=4.5cm,
  %      y radius =3.5cm
   % ] ;
\coordinate (e4) at ($(0,0)+(157:4.5 and 3.5)$);
\draw[thick] (e4) arc
    [
        start angle=157,
        end angle=175,
        x radius=4.5cm,
        y radius =3.5cm
    ] ;

\coordinate (e5) at ($(0,0)+(230:4.5 and 3.5)$);
\coordinate (e6) at ($(0,0)+(225:4.5 and 3.5)$);
\coordinate (e7) at ($(0,0)+(209:4.5 and 3.5)$);
\draw[thick] (e7) arc
    [
        start angle=209,
        end angle=225,
        x radius=4.5cm,
        y radius =3.5cm
    ] ;
\coordinate (h2) at ($(0,0)+(217:4.5*1.1 and 3.5*1.1)$);
\filldraw[black] (h2) circle (1.5pt);
\coordinate (e8) at ($(0,0)+(203:4.5 and 3.5)$);
\draw[thick] (e8) arc
    [
        start angle=203,
        end angle=185,
        x radius=4.5cm,
        y radius =3.5cm
    ] ;

%right elipse
\coordinate (f1) at ($(0,0)+(50:4.5 and 3.5)$);
\coordinate (f2) at ($(0,0)+(45:4.5 and 3.5)$);
\draw[thick] (f2) arc
    [
        start angle=45,
        end angle=29,
        x radius=4.5cm,
        y radius =3.5cm
    ] ;
\coordinate (h3) at ($(0,0)+(37:4.5 and 3.5)$);
\filldraw[black] (h3) circle (1.5pt);
\coordinate (f3) at ($(0,0)+(29:4.5 and 3.5)$);
\coordinate (f4) at ($(0,0)+(23:4.5 and 3.5)$);
\draw[thick] (f4) arc
    [
        start angle=23,
        end angle=5,
        x radius=4.5cm,
        y radius =3.5cm
    ] ;

\coordinate (f5) at ($(0,0)+(310:4.5 and 3.5)$);
\coordinate (f6) at ($(0,0)+(315:4.5 and 3.5)$);
\coordinate (f7) at ($(0,0)+(331:4.5 and 3.5)$);
\draw[thick] (f7) arc
    [
        start angle=331,
        end angle=315,
        x radius=4.5cm,
        y radius =3.5cm
    ] ;
\coordinate (h4) at ($(0,0)+(323:4.5*1.1 and 3.5*1.1)$);
\filldraw[black] (h4) circle (1.5pt);
\coordinate (f8) at ($(0,0)+(337:4.5 and 3.5)$);
\draw[thick] (f8) arc
    [
        start angle=337,
        end angle=355,
        x radius=4.5cm,
        y radius =3.5cm
    ] ;
    
\draw[thick] (e1) arc
    [
        start angle=130,
        end angle=50,
        x radius=4.5cm,
        y radius =3.5cm
    ];
\coordinate (g1) at ($(0,0)+(90:4.5*1.1 and 3.5*1.1)$);
\filldraw[black] (g1) circle (1.5pt);
\draw[thick] (e5) arc
    [
        start angle=230,
        end angle=310,
        x radius=4.5cm,
        y radius =3.5cm
    ] ;
\coordinate (g2) at ($(0,0)+(270:4.5*1.1 and 3.5*1.1)$);
\filldraw[black] (g2) circle (1.5pt);

\draw[dashed,thick] (g1) arc
    [
        start angle=90,
        end angle=175,
        x radius=4.5*1.1cm,
        y radius =3.5*1.1cm
    ] ;
\draw[dashed,thick] (g1) arc
    [
        start angle=90,
        end angle=5,
        x radius=4.5*1.1cm,
        y radius =3.5*1.1cm
    ] ;
\draw[dashed,thick] (g2) arc
    [
        start angle=270,
        end angle=355,
        x radius=4.5*1.1cm,
        y radius =3.5*1.1cm
    ] ;
\draw[dashed,thick] (g2) arc
    [
        start angle=270,
        end angle=185,
        x radius=4.5*1.1cm,
        y radius =3.5*1.1cm
    ] ;

%propagators
\draw[thick] (a1) .. controls (-0.5,0.3) and (0.5,0.3) .. (b1);
\draw[thick] (a2) .. controls (-0.5,0.6) and (0.5,0.6) .. (b2);

\draw[thick] (a3) .. controls (-0.8,1.2) and (0.8,1.2) .. (b3);
\draw[thick] (a4) .. controls (-0.8,1.5) and (0.8,1.5) .. (b4);

\draw[thick] (ad3) .. controls (-0.8,-1.2) and (0.8,-1.2) .. (bd3);
\draw[thick] (ad4) .. controls (-0.8,-1.5) and (0.8,-1.5) .. (bd4);

\draw[thick] (a5) -- (e1);
\draw[thick] (a6) -- (e2);
\draw[thick] (a7) -- (e3);
\draw[thick] (a8) -- (e4);

\draw[thick] (ad5) -- (e5);
\draw[thick] (ad6) -- (e6);
\draw[thick] (ad7) -- (e7);
\draw[thick] (ad8) -- (e8);

\draw[thick] (b5) -- (f1);
\draw[thick] (b6) -- (f2);
\draw[thick] (b7) -- (f3);
\draw[thick] (b8) -- (f4);

\draw[thick] (bd5) -- (f5);
\draw[thick] (bd6) -- (f6);
\draw[thick] (bd7) -- (f7);
\draw[thick] (bd8) -- (f8);

\node[text width=1] at (-0.1,0.85) {$\alpha$};
\node[text width=0.1cm] at (0,-0.3) {$\vdots$};
\node[text width=0.1cm] at (-3.8,0) {$\vdots$};
\node[text width=0.1cm] at (3.8,0) {$\vdots$};

\node[text width=1cm] at (-2,0) {$\mathrm{Tr} Z'^m$};
\node[text width=1cm] at (2,0) {$\mathrm{Tr} Z^n$};
\node[text width=1cm,right,scale=0.8] at ($(a4)+(0.1,0.05)$) {$Z'$};
\node[text width=1cm,scale=0.8] at ($(b4)+(-0.05,0.05)$) {$Z$};
\node[text width=1cm,scale=0.8] at (a7) {$Z'$};
\node[text width=1cm,scale=0.8] at ($(b7)+(0.5,-0.02)$) {$Z$};
\node[text width=1cm,scale=0.8] at ($(e3)+(0.25,-0.25)$) {$X$};
\node[text width=1cm,scale=0.8] at ($(e1)+(0.2,-0.2)$) {$Y$};
\node[text width=1cm,scale=0.8] at ($(f1)+(0.3,-0.25)$) {$X$};
\node[text width=1cm,scale=0.8] at ($(f3)+(0.25,-0.28)$) {$Y$};

%fermions
\node[text width=1cm,scale=0.8] at ($(g1)+(-1.4,-0.45)$) {$\psi^1$} ;
\node[text width=1cm,scale=0.8] at ($(g1)+(+1.4,-0.3)$) {$\bar\psi_2$} ;
\node[text width=1cm,scale=0.8] at ($(e1)+(-0.37,-0.25)$) {$\bar\psi_1$} ;
\node[text width=1cm,scale=0.8] at ($(f1)+(0.82,-0.23)$) {$\psi^2$} ;

\end{tikzpicture}
\caption{The class of ribbon diagrams which contribute to the calculation in the main text at the order of $N^1$.}
\label{fig:diag}
\end{figure}
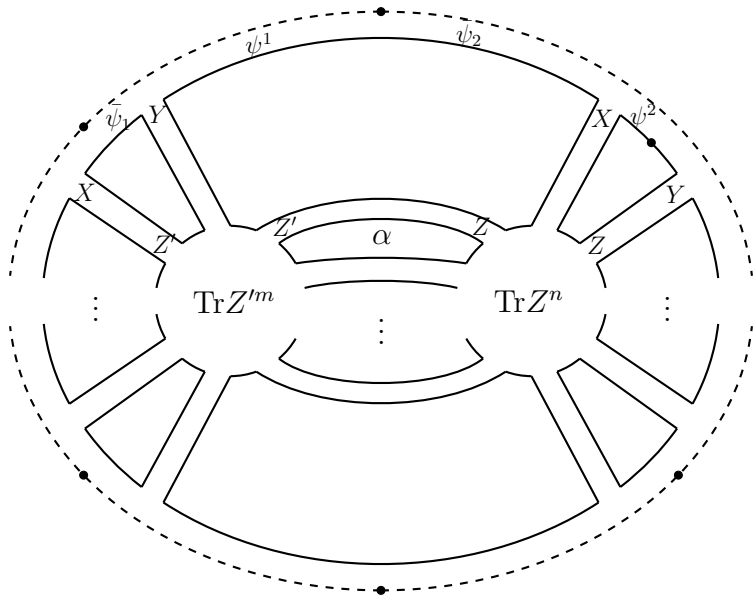

%% file: Det_v4.bbl
\providecommand{\href}[2]{#2}\begingroup\raggedright\begin{thebibliography}{10}

\bibitem{Gopakumar:1998ki}
R.~Gopakumar and C.~Vafa, \emph{{On the gauge theory / geometry
  correspondence}},
  \href{https://doi.org/10.4310/ATMP.1999.v3.n5.a5}{\emph{Adv. Theor. Math.
  Phys.} {\bfseries 3} (1999) 1415}
  [\href{https://arxiv.org/abs/hep-th/9811131}{{\ttfamily hep-th/9811131}}].

\bibitem{Costello:2018zrm}
K.~Costello and D.~Gaiotto, \emph{{Twisted Holography}},
  \href{https://arxiv.org/abs/1812.09257}{{\ttfamily 1812.09257}}.

\bibitem{Bershadsky:1993ta}
M.~Bershadsky, S.~Cecotti, H.~Ooguri and C.~Vafa, \emph{{Holomorphic anomalies
  in topological field theories}},
  \href{https://doi.org/10.1016/0550-3213(93)90548-4}{\emph{Nucl. Phys. B}
  {\bfseries 405} (1993) 279}
  [\href{https://arxiv.org/abs/hep-th/9302103}{{\ttfamily hep-th/9302103}}].

\bibitem{Bershadsky:1993cx}
M.~Bershadsky, S.~Cecotti, H.~Ooguri and C.~Vafa, \emph{{Kodaira-Spencer theory
  of gravity and exact results for quantum string amplitudes}},
  \href{https://doi.org/10.1007/BF02099774}{\emph{Commun. Math. Phys.}
  {\bfseries 165} (1994) 311}
  [\href{https://arxiv.org/abs/hep-th/9309140}{{\ttfamily hep-th/9309140}}].

\bibitem{Jiang:2019xdz}
Y.~Jiang, S.~Komatsu and E.~Vescovi, \emph{{Structure constants in $
  \mathcal{N} $ = 4 SYM at finite coupling as worldsheet g-function}},
  \href{https://doi.org/10.1007/JHEP07(2020)037}{\emph{JHEP} {\bfseries 07}
  (2020) 037} [\href{https://arxiv.org/abs/1906.07733}{{\ttfamily
  1906.07733}}].

\bibitem{Maldacena:1997re}
J.M.~Maldacena, \emph{{The Large N limit of superconformal field theories and
  supergravity}}, \href{https://doi.org/10.1023/A:1026654312961}{\emph{Adv.
  Theor. Math. Phys.} {\bfseries 2} (1998) 231}
  [\href{https://arxiv.org/abs/hep-th/9711200}{{\ttfamily hep-th/9711200}}].

\bibitem{Witten:1998qj}
E.~Witten, \emph{{Anti-de Sitter space and holography}},
  \href{https://doi.org/10.4310/ATMP.1998.v2.n2.a2}{\emph{Adv. Theor. Math.
  Phys.} {\bfseries 2} (1998) 253}
  [\href{https://arxiv.org/abs/hep-th/9802150}{{\ttfamily hep-th/9802150}}].

\bibitem{Beem:2013sza}
C.~Beem, M.~Lemos, P.~Liendo, W.~Peelaers, L.~Rastelli and B.C.~van Rees,
  \emph{{Infinite Chiral Symmetry in Four Dimensions}},
  \href{https://doi.org/10.1007/s00220-014-2272-x}{\emph{Commun. Math. Phys.}
  {\bfseries 336} (2015) 1359}
  [\href{https://arxiv.org/abs/1312.5344}{{\ttfamily 1312.5344}}].

\bibitem{Dedushenko:2019yiw}
M.~Dedushenko and M.~Fluder, \emph{{Chiral Algebra, Localization, Modularity,
  Surface defects, And All That}},
  \href{https://doi.org/10.1063/5.0002661}{\emph{J. Math. Phys.} {\bfseries 61}
  (2020) 092302} [\href{https://arxiv.org/abs/1904.02704}{{\ttfamily
  1904.02704}}].

\bibitem{Pan:2019bor}
Y.~Pan and W.~Peelaers, \emph{{Schur correlation functions on $S^3\times
  S^1$}}, \href{https://doi.org/10.1007/JHEP07(2019)013}{\emph{JHEP} {\bfseries
  07} (2019) 013} [\href{https://arxiv.org/abs/1903.03623}{{\ttfamily
  1903.03623}}].

\bibitem{Antoniadis:1993ze}
I.~Antoniadis, E.~Gava, K.S.~Narain and T.R.~Taylor, \emph{{Topological
  amplitudes in string theory}},
  \href{https://doi.org/10.1016/0550-3213(94)90617-3}{\emph{Nucl. Phys. B}
  {\bfseries 413} (1994) 162}
  [\href{https://arxiv.org/abs/hep-th/9307158}{{\ttfamily hep-th/9307158}}].

\bibitem{Ooguri:1999bv}
H.~Ooguri and C.~Vafa, \emph{{Knot invariants and topological strings}},
  \href{https://doi.org/10.1016/S0550-3213(00)00118-8}{\emph{Nucl. Phys. B}
  {\bfseries 577} (2000) 419}
  [\href{https://arxiv.org/abs/hep-th/9912123}{{\ttfamily hep-th/9912123}}].

\bibitem{Costello:2016mgj}
K.~Costello and S.~Li, \emph{{Twisted supergravity and its quantization}},
  \href{https://arxiv.org/abs/1606.00365}{{\ttfamily 1606.00365}}.

\bibitem{McGreevy:2000cw}
J.~McGreevy, L.~Susskind and N.~Toumbas, \emph{{Invasion of the giant gravitons
  from Anti-de Sitter space}},
  \href{https://doi.org/10.1088/1126-6708/2000/06/008}{\emph{JHEP} {\bfseries
  06} (2000) 008} [\href{https://arxiv.org/abs/hep-th/0003075}{{\ttfamily
  hep-th/0003075}}].

\bibitem{Balasubramanian:2001nh}
V.~Balasubramanian, M.~Berkooz, A.~Naqvi and M.J.~Strassler, \emph{{Giant
  gravitons in conformal field theory}},
  \href{https://doi.org/10.1088/1126-6708/2002/04/034}{\emph{JHEP} {\bfseries
  04} (2002) 034} [\href{https://arxiv.org/abs/hep-th/0107119}{{\ttfamily
  hep-th/0107119}}].

\bibitem{Corley:2001zk}
S.~Corley, A.~Jevicki and S.~Ramgoolam, \emph{{Exact correlators of giant
  gravitons from dual N=4 SYM theory}},
  \href{https://doi.org/10.4310/ATMP.2001.v5.n4.a6}{\emph{Adv. Theor. Math.
  Phys.} {\bfseries 5} (2002) 809}
  [\href{https://arxiv.org/abs/hep-th/0111222}{{\ttfamily hep-th/0111222}}].

\bibitem{Berenstein:2004kk}
D.~Berenstein, \emph{{A Toy model for the AdS / CFT correspondence}},
  \href{https://doi.org/10.1088/1126-6708/2004/07/018}{\emph{JHEP} {\bfseries
  07} (2004) 018} [\href{https://arxiv.org/abs/hep-th/0403110}{{\ttfamily
  hep-th/0403110}}].

\bibitem{Mikhailov:2000ya}
A.~Mikhailov, \emph{{Giant gravitons from holomorphic surfaces}},
  \href{https://doi.org/10.1088/1126-6708/2000/11/027}{\emph{JHEP} {\bfseries
  11} (2000) 027} [\href{https://arxiv.org/abs/hep-th/0010206}{{\ttfamily
  hep-th/0010206}}].

\bibitem{Balasubramanian:2002sa}
V.~Balasubramanian, M.-x.~Huang, T.S.~Levi and A.~Naqvi, \emph{{Open strings
  from N=4 superYang-Mills}},
  \href{https://doi.org/10.1088/1126-6708/2002/08/037}{\emph{JHEP} {\bfseries
  08} (2002) 037} [\href{https://arxiv.org/abs/hep-th/0204196}{{\ttfamily
  hep-th/0204196}}].

\bibitem{Berenstein:2003ah}
D.~Berenstein, \emph{{Shape and holography: Studies of dual operators to giant
  gravitons}},
  \href{https://doi.org/10.1016/j.nuclphysb.2003.10.004}{\emph{Nucl. Phys. B}
  {\bfseries 675} (2003) 179}
  [\href{https://arxiv.org/abs/hep-th/0306090}{{\ttfamily hep-th/0306090}}].

\bibitem{Balasubramanian:2004nb}
V.~Balasubramanian, D.~Berenstein, B.~Feng and M.-x.~Huang, \emph{{D-branes in
  Yang-Mills theory and emergent gauge symmetry}},
  \href{https://doi.org/10.1088/1126-6708/2005/03/006}{\emph{JHEP} {\bfseries
  03} (2005) 006} [\href{https://arxiv.org/abs/hep-th/0411205}{{\ttfamily
  hep-th/0411205}}].

\bibitem{Berenstein:2005vf}
D.~Berenstein and S.E.~Vazquez, \emph{{Integrable open spin chains from giant
  gravitons}}, \href{https://doi.org/10.1088/1126-6708/2005/06/059}{\emph{JHEP}
  {\bfseries 06} (2005) 059}
  [\href{https://arxiv.org/abs/hep-th/0501078}{{\ttfamily hep-th/0501078}}].

\bibitem{Gopakumar:2003ns}
R.~Gopakumar, \emph{{From free fields to AdS}},
  \href{https://doi.org/10.1103/PhysRevD.70.025009}{\emph{Phys. Rev. D}
  {\bfseries 70} (2004) 025009}
  [\href{https://arxiv.org/abs/hep-th/0308184}{{\ttfamily hep-th/0308184}}].

\bibitem{Gopakumar:2004qb}
R.~Gopakumar, \emph{{From free fields to AdS. 2.}},
  \href{https://doi.org/10.1103/PhysRevD.70.025010}{\emph{Phys. Rev. D}
  {\bfseries 70} (2004) 025010}
  [\href{https://arxiv.org/abs/hep-th/0402063}{{\ttfamily hep-th/0402063}}].

\bibitem{Gopakumar:2004ys}
R.~Gopakumar, \emph{{Free field theory as a string theory?}},
  \href{https://doi.org/10.1016/j.crhy.2004.10.004}{\emph{Comptes Rendus
  Physique} {\bfseries 5} (2004) 1111}
  [\href{https://arxiv.org/abs/hep-th/0409233}{{\ttfamily hep-th/0409233}}].

\bibitem{Gopakumar:2005fx}
R.~Gopakumar, \emph{{From free fields to AdS: III}},
  \href{https://doi.org/10.1103/PhysRevD.72.066008}{\emph{Phys. Rev. D}
  {\bfseries 72} (2005) 066008}
  [\href{https://arxiv.org/abs/hep-th/0504229}{{\ttfamily hep-th/0504229}}].

\bibitem{Aharony:2007fs}
O.~Aharony, J.R.~David, R.~Gopakumar, Z.~Komargodski and S.S.~Razamat,
  \emph{{Comments on worldsheet theories dual to free large N gauge theories}},
  \href{https://doi.org/10.1103/PhysRevD.75.106006}{\emph{Phys. Rev. D}
  {\bfseries 75} (2007) 106006}
  [\href{https://arxiv.org/abs/hep-th/0703141}{{\ttfamily hep-th/0703141}}].

\bibitem{Gaberdiel:2021iil}
M.R.~Gaberdiel and R.~Gopakumar, \emph{{The String Dual to Free ${\cal N}=4$
  Super Yang-Mills}},  \href{https://arxiv.org/abs/2104.08263}{{\ttfamily
  2104.08263}}.

\bibitem{Gaberdiel:2021jrv}
M.R.~Gaberdiel and R.~Gopakumar, \emph{{The Worldsheet Dual of Free Super
  Yang-Mills in 4D}},  \href{https://arxiv.org/abs/2105.10496}{{\ttfamily
  2105.10496}}.

\bibitem{Gopakumar:2011ev}
R.~Gopakumar, \emph{{What is the Simplest Gauge-String Duality?}},
  \href{https://arxiv.org/abs/1104.2386}{{\ttfamily 1104.2386}}.

\bibitem{Gopakumar:2012ny}
R.~Gopakumar and R.~Pius, \emph{{Correlators in the Simplest Gauge-String
  Duality}}, \href{https://doi.org/10.1007/JHEP03(2013)175}{\emph{JHEP}
  {\bfseries 03} (2013) 175} [\href{https://arxiv.org/abs/1212.1236}{{\ttfamily
  1212.1236}}].

\bibitem{Lin:2004nb}
H.~Lin, O.~Lunin and J.M.~Maldacena, \emph{{Bubbling AdS space and 1/2 BPS
  geometries}},
  \href{https://doi.org/10.1088/1126-6708/2004/10/025}{\emph{JHEP} {\bfseries
  10} (2004) 025} [\href{https://arxiv.org/abs/hep-th/0409174}{{\ttfamily
  hep-th/0409174}}].

\bibitem{Gaiotto:2009gz}
D.~Gaiotto and J.~Maldacena, \emph{{The Gravity duals of N=2 superconformal
  field theories}}, \href{https://doi.org/10.1007/JHEP10(2012)189}{\emph{JHEP}
  {\bfseries 10} (2012) 189} [\href{https://arxiv.org/abs/0904.4466}{{\ttfamily
  0904.4466}}].

\bibitem{indet}
D.~Gaiotto and J.H.~Lee, \emph{The giant graviton expansion}, {\emph{To appear}
  }.

\bibitem{Dijkgraaf:2002dh}
R.~Dijkgraaf and C.~Vafa, \emph{{A Perturbative window into nonperturbative
  physics}},  \href{https://arxiv.org/abs/hep-th/0208048}{{\ttfamily
  hep-th/0208048}}.

\bibitem{Dijkgraaf:2002fc}
R.~Dijkgraaf and C.~Vafa, \emph{{Matrix models, topological strings, and
  supersymmetric gauge theories}},
  \href{https://doi.org/10.1016/S0550-3213(02)00766-6}{\emph{Nucl. Phys. B}
  {\bfseries 644} (2002) 3}
  [\href{https://arxiv.org/abs/hep-th/0206255}{{\ttfamily hep-th/0206255}}].

\bibitem{Aganagic:2003qj}
M.~Aganagic, R.~Dijkgraaf, A.~Klemm, M.~Marino and C.~Vafa, \emph{{Topological
  strings and integrable hierarchies}},
  \href{https://doi.org/10.1007/s00220-005-1448-9}{\emph{Commun. Math. Phys.}
  {\bfseries 261} (2006) 451}
  [\href{https://arxiv.org/abs/hep-th/0312085}{{\ttfamily hep-th/0312085}}].

\end{thebibliography}\endgroup
